%% file: main.tex
\documentclass[10pt,twocolumn,letterpaper]{article}

\usepackage[pagenumbers]{cvpr} 

\input{preamble}

\definecolor{cvprblue}{rgb}{0.21,0.49,0.74}
\usepackage[pagebackref,breaklinks,colorlinks,allcolors=cvprblue]{hyperref}

\def\confName{CVPR}
\def\confYear{2027}

\title{Learning the Next Best Move with Visual Diffusion Models}

\author{
{\fontsize{11pt}{13pt}\selectfont Amir Razlighi$^{1,2,3}$\quad Ali Mahdavi Amiri$^{1}$\quad Yiorgos Chrysanthou$^{2,3}$\quad Andrey Voynov$^{4}$\quad Daniel Cohen-Or$^{5}$}\\[6pt]
{\small $^{1}$Simon Fraser University\quad $^{2}$University of Cyprus\quad $^{3}$CYENS Centre of Excellence\quad $^{4}$Google DeepMind\quad $^{5}$Tel Aviv University}
}

\begin{document}
\maketitle
\begin{strip}
    \centering
    \includegraphics[width=\textwidth]{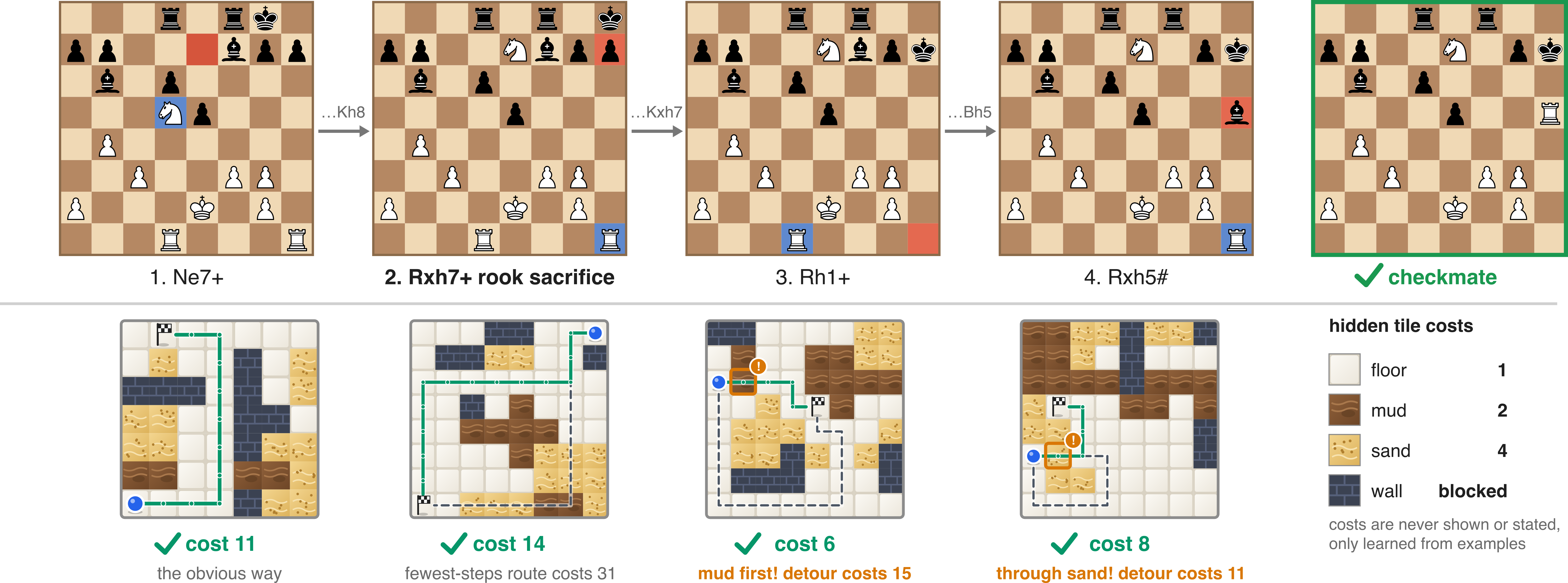}
    \captionof{figure}{
    Visual reasoning without explicit lookahead.
Our model solves a chess puzzle one move at a time (top), including a
rook sacrifice leading to checkmate, and finds the cheapest route across
terrain whose costs are never shown (bottom), directly from images.
Blue and red mark each chess move's source and destination.
}
    \label{fig:teaser}
\end{strip}

\input{sec/0_abstract}    
\input{sec/1_intro}

\input{sec/Y_examples}
\input{sec/3_method}
\input{sec/Analysis}
\input{sec/late-related}
\input{sec/conclusion}

{
    \small
    \bibliographystyle{ieeenat_fullname}
    \bibliography{main}
}

\input{sec/X_suppl}

\end{document}

%% file: preamble.tex
\usepackage{float}
\usepackage[table]{xcolor}
\usepackage{multirow}
\usepackage{booktabs}
\definecolor{saffron}{rgb}{0.95,0.75,0.2}

\usepackage{cuted}

%% file: sec/0_abstract.tex
\begin{abstract}
In this paper, we show that visual diffusion models can learn to predict the next best move (NBM) in tasks requiring sequential reasoning and lookahead, without explicitly exploring future steps or exhaustive searching at inference. Previous work demonstrated that these models can generate complete solutions to geometric and reasoning problems.
Here, we show that simple fine-tuning on visual state--NBM pairs enables a pretrained model to select actions whose value depends on consequences several steps ahead. Each move is predicted directly in image space from the current state, without an explicit symbolic representation or search procedure. Repeated NBM predictions then construct a solution one decision at a time.

We demonstrate this capability on chess puzzles and tile-based pathfinding, and show that it extends to Connect Four, Go and billiards. In chess, the model predicts legal moves and finds the best move far more often than by recalling similar training positions, including piece sacrifices whose benefit emerges only through later tactical continuations. In pathfinding, it constructs shortest paths on uniform-cost maps and cost-optimal routes across terrain tiles with different traversal costs, choosing longer routes when they are cheaper overall. These costs are learned implicitly from examples, without numerical cost supervision.
These results show that direct visual move prediction
can support globally informed decisions and successful sequential problem solving.
\end{abstract}

%% file: sec/1_intro.tex
\section{Introduction}
\label{sec:intro}

\noindent\textit{``I see only one move ahead, but it is always the correct one.''}
--- \textit{attributed to Jos\'e Ra\'ul Capablanca}
\vspace{0.4em}

Visual generative models have demonstrated a surprising ability to solve
hard problems by generating their solutions as images. Beyond synthesizing
realistic appearance, visual diffusion models can produce geometric
configurations, complete puzzles, and construct paths that satisfy complex
constraints~\cite{goren2025visual,wewer2025spatial,he2025diffthinker,ezquerro2026relational}.
These findings suggest that visual generation can support forms of
reasoning traditionally addressed by specialized algorithms. In this work,
we investigate whether visual diffusion models can also learn to predict
the next best move (NBM): choosing an action that advances a task toward
its solution, without searching over future states. It's worth mentioning that our goal is not to develop a task-specific solver (e.g., a competitive chess player), but to demonstrate and analyze an intriguing capability of visual diffusion models. We use chess, pathfinding, and other tasks as testbeds to investigate and better understand this capability.

Choosing the next move introduces a further challenge: its quality
depends on what can happen afterward. A valid action needs only to satisfy
the rules of the current state; the best action must also account for
consequences several steps ahead. Generating a complete solution allows
its parts to be represented and refined together. NBM prediction instead
commits to a single action without producing the continuation that
justifies it. The challenge is therefore to capture the benefit of
lookahead through direct visual prediction, without explicitly exploring
future steps at inference.

Chess puzzles provide a concrete setting for this challenge.
From a board image alone, the model must predict a move that is
both legal and directed toward solving the puzzle against the
opponent's best responses. Piece sacrifices are particularly
striking: their benefit emerges only through later consequences.
Section~\ref{sec:reasoning-in-action} develops this example alongside
cost-aware pathfinding, illustrating the decisions we seek to learn.

We show that a pretrained visual diffusion model can learn to make
such decisions by fine-tuning on pairs of visual states and their
next best moves. Given the current state as an image, the model
directly generates a visual prediction of the NBM, without explicit
search or symbolic reasoning modules. After the move is applied,
the updated state serves as input for the next prediction.
Repeated application constructs a solution one decision at a time,
selecting actions whose benefit depends on later consequences.
The same formulation extends to predicting a short sequence of
future moves at once. The number of moves predicted is commonly
referred to as the prediction horizon. A video diffusion model
marks successive moves and generates their resulting states,
alternating between the solver and the opponent in chess.

In our work, we show that, under the same training budget, predicting several moves together
improves the accuracy of the next move, even when the remaining
predictions are not executed. Generating a continuation therefore
offers a benefit beyond producing a single action. In deterministic
tasks, the model can also execute several predicted moves before
being called again. Against an opponent, whose replies may differ
from those predicted, we execute only the first move and make
a new prediction after observing the reply.



Our study of next best move (NBM) prediction reveals a surprising finding: predicting the move from the first denoising step is more accurate than using a fully denoised sample. At this early stage, several possible moves appear as overlapping marks in a single image. The strongest marks often reveal the correct move. As denoising proceeds, the model settles on one candidate, which may differ from the move most strongly indicated at the first step. We can thus read out the model’s preferred move before denoising is complete.

%% file: sec/Y_examples.tex
\section{Visual Reasoning in Action}
\label{sec:reasoning-in-action}

We first illustrate, on chess puzzles and cost-aware pathfinding, what the model observes, how it expresses an action, and why choosing that action requires accounting for later consequences.

\begin{figure}[t]
    \centering
    \includegraphics[width=\linewidth]{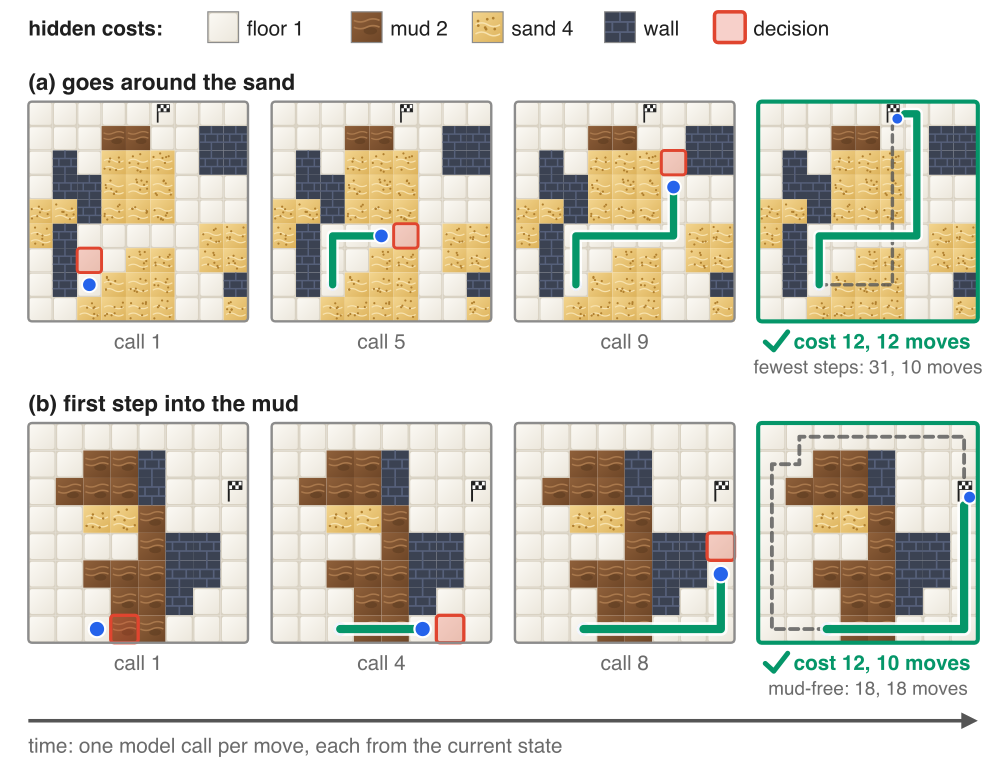}
    \caption{Cost-aware navigation through successive next best move (NBM)
predictions on two held-out maps. \emph{(a)} The model chooses longer routes when they reduce total traversal cost by avoiding expensive terrain, and \emph{(b)} accepts some high cost actions in cases where it makes the path much shorter. Terrain costs are learned implicitly from example solutions, without explicit numerical supervision.}
    \label{fig:terrain-paths}
\end{figure}

\begin{figure}[t]
    \centering
    \includegraphics[width=\linewidth]{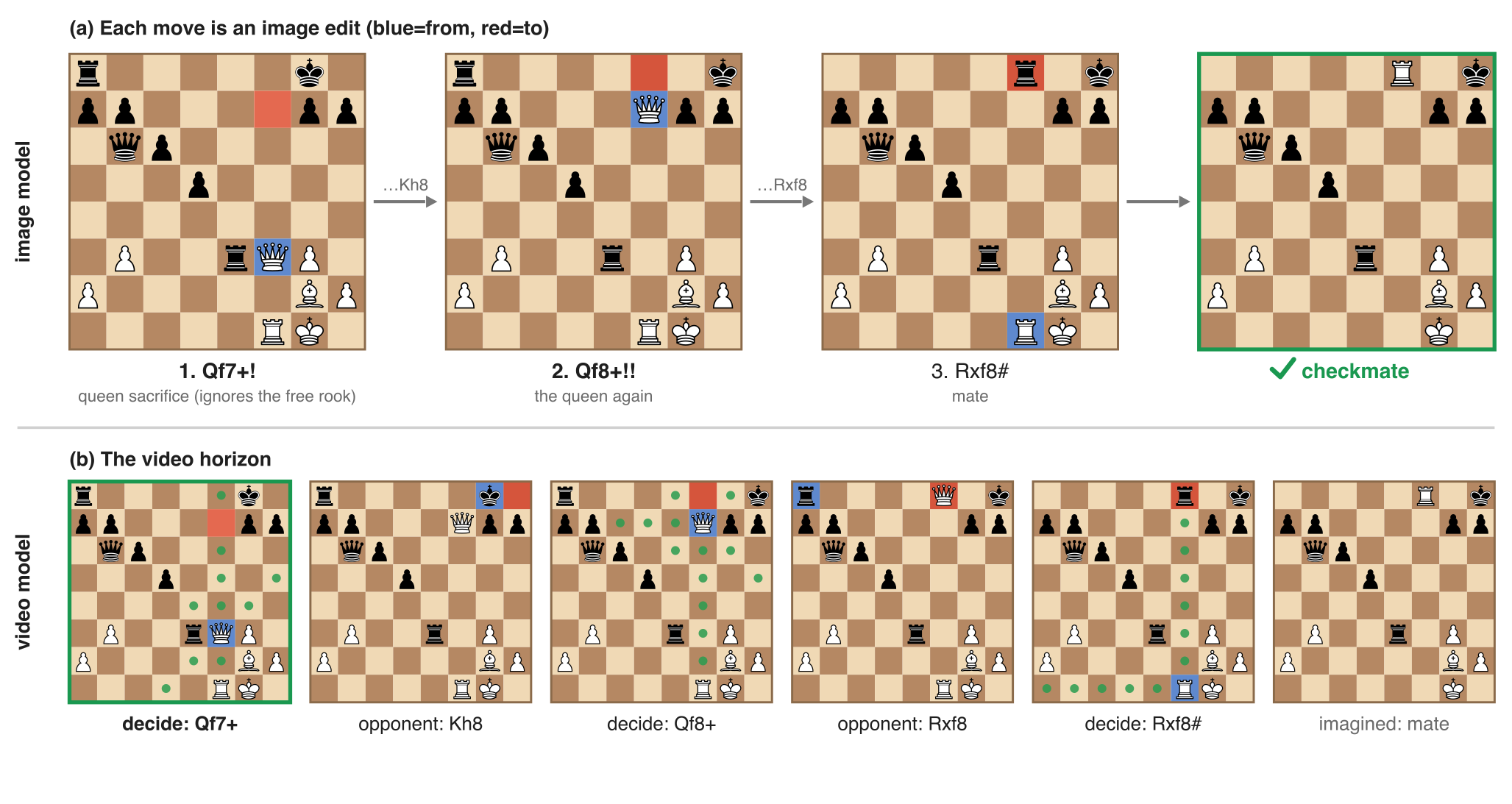}
    \caption{A double queen sacrifice found without search, \emph{(a)} by
    the image model, which predicts one move per call on the current board,
    and \emph{(b)} by the video model, which predicts several moves together
    with the board states they lead to.}
    \label{fig:chess-sacrifice}
\end{figure}

\subsection{Chess Puzzles}

A chess puzzle presents a board position and asks the player to find
a sequence of moves that achieves a tactical objective, such as
checkmate or a decisive material advantage. Solving it requires
choosing moves that succeed against the opponent's best responses.
Many moves may be legal, but only a few advance toward the solution.

Our model receives the board as an image and predicts a move by
marking its source and destination squares. It receives no symbolic
board encoding, piece identities, or explicit movement rules.
A bishop, for example, is represented only by its visual appearance;
the model must learn from visual examples how it moves and when
moving it is useful. Fine-tuning pairs show board images and their
next best moves, providing supervision for both valid movement
patterns and effective action selection.

Figure~\ref{fig:chess-sacrifice} presents a striking example of a decision whose value becomes apparent only through its later consequences. In this held-out puzzle, White can capture an undefended rook with Qxe3, gaining material immediately. Instead, the model predicts Qf7+, offering its queen to force the black king into the corner. After the reply Kh8 is applied and the new board is fed back, it predicts Qf8+, sacrificing the queen a second time to draw the black rook onto f8, and after Black's Rxf8 it delivers mate with Rxf8\#. The sacrifice is particularly revealing as it gives up the most valuable piece despite an immediate material gain.

Each move is predicted separately from the current board.
The model commits to the queen sacrifice before generating any
subsequent move, without constructing future board states,
evaluating alternative continuations, or performing explicit tree
search. Successive NBM predictions
recover the displayed tactical sequence, with each prediction
matching the puzzle's prescribed move. This example illustrates
how direct visual prediction can select an action whose
justification depends on a coordinated continuation several
moves ahead, without explicitly generating that continuation
at inference. The example is not an isolated case: on the 1{,}001 held-out puzzles tagged as sacrifices, the video model finds the next best move 47.9\% of the
time (57.3\% with voting), against 16.4\% for copying the move of the nearest training position (\cref{tab:chess_recall_vs_play}).

\subsection{Cost-Aware Pathfinding}

Figure~\ref{fig:terrain-paths} shows two examples of navigation on
held-out tile maps. The agent moves between adjacent traversable tiles,
and each terrain type has a fixed cost for entering a tile. The objective
is to reach the goal while minimizing the sum of tile costs along the
route. With uniform tile costs, this reduces to finding a shortest path;
with terrain-dependent tile costs, a longer route can be cheaper.
Both settings can be solved by Dijkstra's algorithm, which explicitly
expands candidate states and maintains cumulative distance estimates.
Our model instead predicts one move at a time, without explicitly
exploring future routes.

The model receives no numerical terrain tile costs during training,
learning only from examples of optimal next moves. Through successive
NBM predictions, it constructs cost-optimal routes, taking additional
steps when these avoid sufficiently expensive tiles. Each move is
predicted from the current visual state, without generating the
remaining route or comparing alternative future continuations.
These examples illustrate decisions whose quality depends on costs
several steps ahead: entering a cheap tile can lead to an expensive
continuation, whereas an initially indirect move can enable a cheaper
route overall. This holds beyond the two examples: on held-out maps where the cheapest
route either detours around or crosses an expensive band, the model makes
the right choice on $85.0$\% of maps against $50$\% for chance, and it finds
the cheapest route on $87.5$\% of test maps against $59.2$\% for the fewest-steps
route (\cref{tab:extended_tasks}).

%% file: sec/3_method.tex
\section{Method}
\label{sec:method}

We turn a pretrained visual generative model into a decision maker without changing its architecture: the state is an image, the action is an edit of that image, and decisions are made by sampling.

\subsection{Next best move prediction}
\label{sec:method_nbm}

A task is a sequence of states $s_0, s_1, \dots$ observed only through
renderings $x_t = R(s_t)$. Taking action $a_t$ in state $s_t$ leads to
$s_{t+1} = E(s_t, a_t)$, where in two-player games, the environment $E$ also applies the
opponent's reply. Training data provide, for each state, a reference
next best move $a^\star_t$ taken from a known solution. We write $M(x,a)$
for the image $x$ with action $a$ drawn on it (\cref{sec:method_encoding})
and train a conditional generator $p_\theta$ to produce $M(x_t,
a^\star_t)$ from $x_t$ and a fixed text prompt $c$. At inference, a
deterministic decoder $D$ reads the action from the generated image,
\begin{equation}
  \hat a_t = D(\hat y_t, x_t), \qquad \hat y_t \sim p_\theta(\,\cdot \mid x_t, c),
  \label{eq:nbm}
\end{equation}
the environment returns $s_{t+1} = E(s_t, \hat a_t)$, and the model is
called again on $x_{t+1}$. A solution is thus built one decision at a
time, in closed loop.

\noindent\textbf{No search.}
Each call draws a single noise sample and runs a fixed number of
denoising steps. The model never enumerates, simulates or scores
alternative actions, and neither the environment nor any engine is
queried within a call. Whatever the decision anticipates about later
steps must therefore be carried by a single generation.

\subsection{Actions as image edits}
\label{sec:method_encoding}

An action moves an entity from cell $u$ to cell $v$ of a grid, such as a chess piece between squares or an agent to a neighbouring tile, or places an entity on a cell $v$, such as a stone in Go. $M(x, a)$ marks $u$ in blue and $v$ in red (only $v$ for placements) and leaves all other pixels unchanged (\cref{fig:chess-sacrifice}), so input and target differ only where the action takes place, and the output space is the image itself rather than a task specific action vocabulary. The prompt is fixed per task and contains only what the image cannot show, such as which side should move and how the predicted move should be marked on the image. Rules, move generators and costs are never provided as input; in pathfinding with terrain, the cost of each tile type is neither drawn nor stated. What the model knows about them is learned from the training targets alone.

\noindent\textbf{Decoding.}
$D$ inverts the marking. For each cell, it fits the difference between
$\hat y_t$ and $x_t$ as a blend toward the blue and the red tint, giving
strengths $\alpha_b(u)$ and $\alpha_r(v)$, and returns the pair of the
strongest blue and the strongest red cell. This readout uses no rules,
and an action that cannot be executed counts as a failure.

\subsection{From one move to a horizon}
\label{sec:method_horizon}

The closed loop of \cref{eq:nbm} makes one NBM prediction per call. A
horizon model makes $H$ of them in a single generation: it predicts the
NBM on the current state, the state that move leads to, the NBM on that
state, and so on. For a reference continuation $a^\star_t, \dots,
a^\star_{t+H-1}$ with visual states $x^\star_{t+k} = R(s^\star_{t+k})$ and
$x^\star_t = x_t$, the target is
\begin{equation}
  \big(\, M(x^\star_{t+k-1}, a^\star_{t+k-1}),\; x^\star_{t+k} \,\big)_{k=1}^{H},
  \label{eq:horizon}
\end{equation}
rendered as a video that starts from $x_t$ and alternates \emph{Decide} frames, the current state with the next move marked, and \emph{Imagine} frames, the visual state after that move. For a single agent, the horizon consists of the agent's successive NBMs; in two-player games, it alternates between the sides, and the NBM predicted for the opponent is its best reply. In chess, the video model's \emph{Decide} frames also mark the moving piece's legal destinations with green dots, which the decoder does not read (Appendix~\ref{app:models}) and which serve to check whether the model has learned how each piece moves. Each move is decoded from the generated frames, $\hat a_{t+k-1} = D(\hat m_k, \hat x_{k-1})$ with $\hat x_0 = x_t$, so every predicted move is an explicit, checkable decision on an explicit state.

\noindent\textbf{Executing a horizon.}
How many predicted moves to execute depends on how much of the future the current state determines. In a deterministic single-agent task, they can be executed directly: this replaces $H$ calls with one, and when $H$ equals the solution length $L$, the entire solution is drawn in a single image. Against an opponent, moves beyond the next depend on replies that may differ from the predicted ones, so we use a receding horizon~\cite{mayne2000constrained}: only the first predicted move is executed, the true state is observed, and the model is called again. This recovers the closed loop of \cref{eq:nbm}, but each decision now comes with the continuation the model expects to follow from it. One formulation thus covers single decisions ($H=1$), decisions with an explicit continuation (receding horizon) and complete solutions ($H=L$); \cref{sec:experiments} compares them.

\subsection{Models and training}
\label{sec:method_models}

We instantiate $p_\theta$ with two pretrained models and adapt both with LoRA~\cite{hu2021lora}, adding no new modules. For $H=1$ and for whole-solution images we use the image-editing model Qwen-Image-2.1~\cite{wu2025qwen}, which conditions on $x_t$ through its multimodal text encoder and through clean image latents placed before the noisy target latents. For $H>1$ we use the text-image-to-video model Wan2.2-TI2V-5B~\cite{wan2025wan}, with $x_t$ as its first frame; \emph{Decide} and \emph{Imagine} frames are generated jointly. Both models keep their rectified-flow objective~\cite{liu2022flow,lipman2022flow}: with target
latents $z_0$, noise $\epsilon \sim \mathcal N(0, I)$ and
$z_\sigma = (1-\sigma)\, z_0 + \sigma \epsilon$,
\begin{equation}
  \mathcal L(\theta) = \mathbb E_{z_0, \epsilon, \sigma}
  \big\| v_\theta(z_\sigma, \sigma, x_t, c) - (\epsilon - z_0) \big\|^2 .
  \label{eq:loss}
\end{equation}

\begin{table*}[t]
\centering
\small
\setlength{\tabcolsep}{2.5pt}
\renewcommand{\arraystretch}{1.05}
\begin{tabular}{@{}l ccc cc ccc@{}}
\toprule
& \multicolumn{3}{c}{NBM Accuracy (\%)}
& \multicolumn{2}{c}{Symmetric boards (0 = no change)}
& \multicolumn{3}{c}{One piece changed (\%)} \\
\cmidrule(lr){2-4} \cmidrule(lr){5-6} \cmidrule(l){7-9}
& All $\uparrow$ & Sacrifices $\uparrow$ & Far from
& Colour & Mirror
& Keeps move: & Keeps move: & Finds new \\
& & & training $\uparrow$
& flip &
& best stays $\uparrow$ & best changes $\downarrow$ & best move $\uparrow$ \\
\midrule
\multicolumn{9}{@{}l}{\textit{Ours}} \\
Image model                      & 56.0 & 46.3 & 50.4 & $+1.2$ & $-1.9$ & 82.3 & \underline{35.3} & 39.5 \\
\quad + majority vote            & \underline{63.8} & \underline{54.9} & \underline{58.7} & $+1.2$ & $-1.5$ & 89.7 & 37.7 & 42.5 \\
\quad + oracle of 8 samples$^\dagger$ & \textit{85.4} & \textit{83.5} & \textit{82.3} & -- & -- & -- & -- & -- \\
\quad + one step                 & 62.5 & 51.7 & 56.6 & \textbf{$-$0.1} & $-1.8$ & 94.7 & 35.9 & 43.6 \\
\quad + one step + majority vote & 63.2 & 53.0 & 57.9 & \textbf{$-$0.1} & $-1.8$ & \underline{94.8} & 36.1 & \underline{44.4} \\
Video model                      & 56.4 & 47.9 & 51.0 & \textbf{$+$0.1} & $-2.7$ & 94.3 & 42.0 & 38.4 \\
\quad + majority vote            & \textbf{65.4} & \textbf{57.3} & \textbf{61.5} & \textbf{$-$0.1} & $-2.9$ & 94.5 & 39.7 & 42.6 \\
\quad + oracle of 8 samples$^\dagger$ & \textit{86.6} & \textit{82.1} & \textit{84.1} & -- & -- & -- & -- & -- \\
\quad + one step                 & 63.9 & 55.6 & 59.1 & $+0.7$ & \underline{$-$1.1} & 95.8 & 37.0 & 45.5 \\
\quad + one step + majority vote & 64.7 & 56.2 & 60.0 & $+0.9$ & \textbf{$-$0.5} & \textbf{95.9} & 36.6 & \textbf{46.6} \\
\midrule
\multicolumn{9}{@{}l}{\textit{Copying from training}} \\
Nearest training position   & 18.6 & 16.4 & \phantom{0}8.7 & $+1.2$ & $-4.3$ & 36.2 & \textbf{32.1} & \phantom{0}6.7 \\
Majority of 16 nearest      & 27.0 & 27.0 & 16.8 & $+0.8$ & $-6.9$ & 56.5 & 48.1 & 12.5 \\
Random legal move           & \phantom{0}5.5 & \phantom{0}3.8 & \phantom{0}4.0 & -- & -- & -- & -- & -- \\
\bottomrule
\end{tabular}
\caption{\textbf{Recall or play?} We compare our models on held-out chess puzzles, with single samples and with majority votes, against copying moves from the most similar training positions, both on held-out puzzles (``far'': 13--16 squares differ from the nearest training position) and on the same puzzles after a symmetry transform or a single-piece change (which can change the NBM dramatically). $^\dagger$At least one of 8 samples is correct. All models are trained on three million puzzles. More details and variants are in Appendix~\ref{app:recall}.
}
\label{tab:chess_recall_vs_play}
\end{table*}

\noindent\textbf{Test-time extensions.}
Since NBM is a standard conditional generation task, existing test-time techniques for diffusion models apply directly. We report results with and without two of them, majority voting and sampling with few denoising steps (\cref{sec:experiments}); neither is required for our main claims or adds an engine, a rule-based module, or search at inference. A third, contrastive guidance away from a learned losing-move prompt, is evaluated in Appendix~\ref{app:guidance}.

\noindent\emph{Majority voting.} The model is sampled $n$ times from the same state with independent noise, and the most frequent decoded move is executed,
\begin{equation}
  \hat a_t = \operatorname*{mode}_{i=1,\dots,n} D\big(\hat y_t^{(i)}, x_t\big),
  \qquad \hat y_t^{(i)} \sim p_\theta(\cdot \mid x_t, c),
  \label{eq:vote}
\end{equation}
same as the self-consistency rule of language models~\cite{wang2022self}: it aggregates the model's own predictions and compares no continuations.

%% file: sec/Analysis.tex
\section{Analysis and Ablations}
\label{sec:experiments}

\noindent\textbf{Setup.}
Chess positions are drawn from the Lichess puzzle database and split by game to prevent overlap between training and evaluation. Our main models are trained on up to three million puzzles and evaluated on 10{,}000 held-out positions (Appendix~\ref{app:task-chess}). The image model predicts the next move of each puzzle solution, while the video model predicts up to eight plies, including the opponent's replies. The remaining tasks are generated with exact solvers or engines, used only to construct training targets and score predictions (Appendix~\ref{app:tasks}). Unless stated otherwise, we use $20$ Euler steps for the image model and $30$ UniPC steps for the video model, and invalid actions count as failures. We report next-best-move accuracy for chess and the fraction of optimal routes for pathfinding.

\begin{figure}[t]
    \centering
    \includegraphics[width=\linewidth]{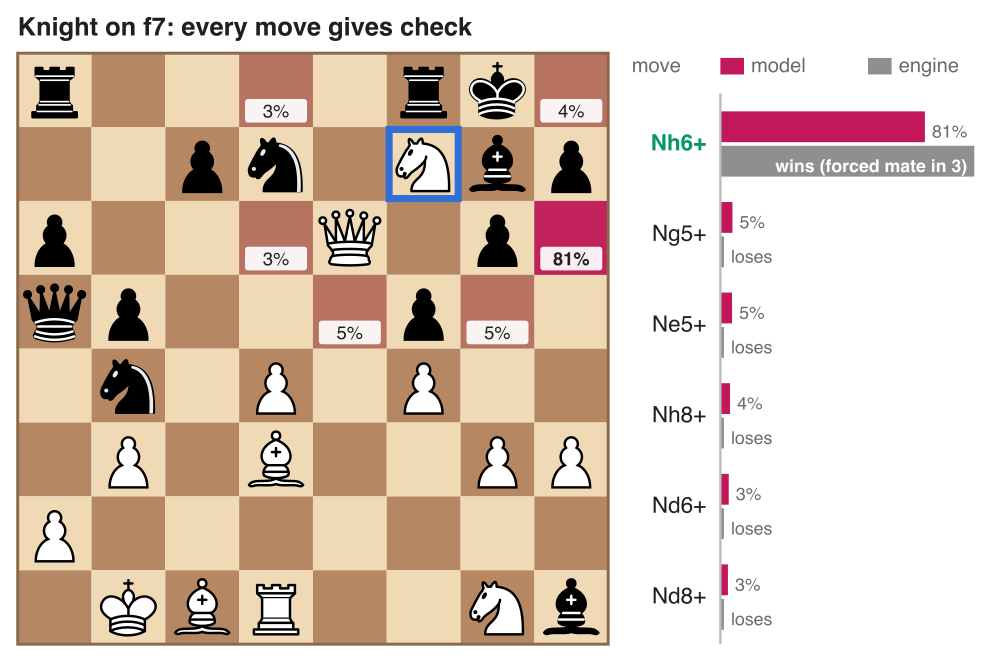}
    \caption{\textbf{Implicit policy.} Every knight move gives check here, but scoring each move by the model's denoising error puts 81\% on Nh6+, the only winning move.}
    \label{fig:implicit-policy}
\end{figure}

\noindent\textbf{Recall or play?}
To test whether these decisions reflect the position rather than recalled
training examples, we compare against retrieval from the training set,
evaluate on boards transformed by exact symmetries of the game, and alter
single pieces so that the best move changes: the model's choices track
these changes, while retrieval does not.

\begin{figure}[t]
    \centering
    \includegraphics[width=\linewidth]{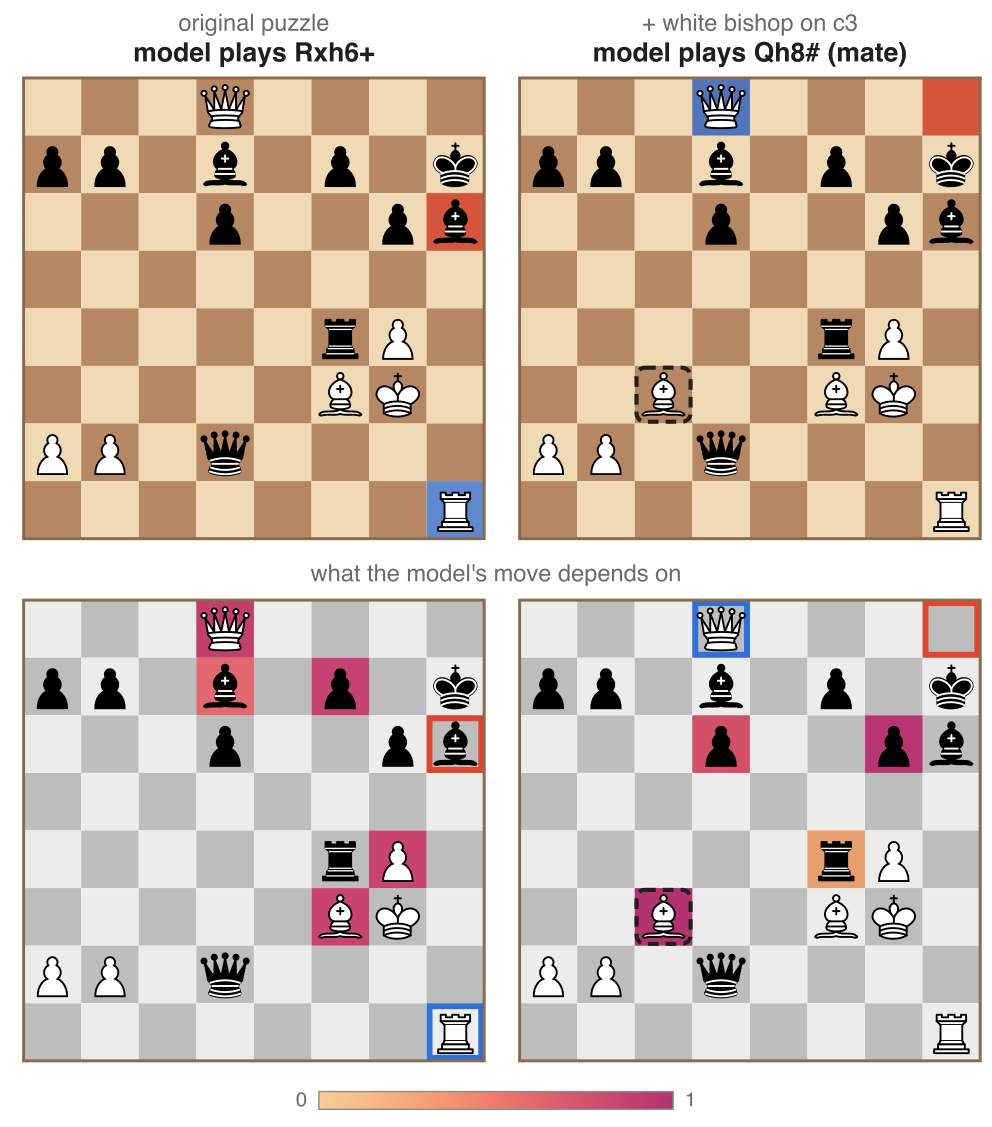}
    \caption{\emph{``top'':} Two similar board configurations, where the only difference is an added white bishop on c3. \emph{``bottom'':}  how much our model's confidence in the move drops when each piece is removed.
    }
    \label{fig:bishop-attention}
\end{figure}

\noindent\textbf{Is the model relying on the right pieces?}
Figure~\ref{fig:bishop-attention} shows the same effect on a single held-out puzzle (rated 1932). On the original board, our model plays the rook sacrifice Rxh6+, which forces mate in three. Adding a white bishop on c3 leaves the position almost unchanged visually, yet it changes the best move: the bishop now guards h8, so Qh8\# mates in one, and the model switches to exactly this move. To see what each decision rests on, we remove one piece at a time and measure how much the model's confidence in its move drops. On the original board, the sacrifice depends on the pieces that build the later mating net, namely the queen that delivers the final checks and the bishop and pawn that cover the king's escape squares. With the added bishop, the decision depends most on the two pieces that make the new mate work: the bishop guarding h8 and Black's own pawn on g6, which blocks the king's only escape. Removing either of them indeed makes Qh8 no longer mate. The model thus separates two boards that differ in a single piece as the position requires, and Table~\ref{tab:chess_recall_vs_play} shows that this holds over thousands of such changes.

\begin{table*}[t]
  \centering
\small
\setlength{\tabcolsep}{10.0pt}
\renewcommand{\arraystretch}{1.10}
\begin{tabular}{@{}l l l cc l c@{}}
\toprule
Task & Action drawn visually & Metric (\%, $\uparrow$) & Image & Video & Strongest baseline & \\
\midrule
\multirow{3}{*}{Shortest path}
  & \multirow{3}{*}{route on a grid}
  & optimal, held-out maps        & 88.8 & \textbf{94.6} & base model, zero-shot & 0.0 \\
  & & optimal, longer paths       & 47.0 & \textbf{66.6} & base model, zero-shot & 0.0 \\
  & & optimal, larger maps        & 79.0 & \textbf{79.8} & base model, zero-shot & 0.0 \\
\midrule
\multirow{2}{*}{\shortstack[l]{Cost-aware path\\(costs never shown)}}
  & \multirow{2}{*}{route on terrain}
  & cheapest route               & \textbf{87.5} & 82.8 & fewest steps (BFS) & 59.2 \\
  & & detour or cross            & \textbf{85.0} & 76.8 & chance & 50.0 \\
\midrule
\multirow{3}{*}{Connect Four}
  & \multirow{3}{*}{disc in a column}
  & legal, correct landing row   & \textbf{100.0} & 99.6 &  & \\
  & & keeps the win w/o lookahead & \textbf{67.5} & 64.5 &  & \\
  & & \quad + one-step            & \textbf{76.4} & 72.0 & one-move heuristic & 60.4 \\
\midrule
\multirow{3}{*}{\shortstack[l]{Go\\($19{\times}19$, KGS)}}
  & \multirow{3}{*}{stone on a point}
  & legal move                   & 99.8 & -- &  &  \\
  & & best move                  & 20.9 & -- & random legal move & 0.7 \\
  & & \quad + one-step           & \textbf{31.1} & -- & copy nearest position & 1.7 \\
\midrule
\multirow{2}{*}{Billiards}
  & \multirow{2}{*}{cue angle, power}
  & ball potted                  & 18.6 & \textbf{34.6} & copy nearest shot & 19.0 \\
  & & \quad + one-step           & 28.0 & \textbf{47.9} & random shot & 0.7 \\
\bottomrule
\end{tabular}
\caption{\textbf{Beyond chess.} The same recipe, unchanged, on tasks whose actions are drawn as a route, a disc, a
stone, or a continuous shot. Rules, tile costs, and physics are never given; the models learn them from example
solutions alone. Routes are built move by move: the image model marks one move per call, and the video model
generates several moves per clip.}
\label{tab:extended_tasks}
\end{table*}

\noindent\textbf{Implicit policy.}
A diffusion model can also score outputs it did not sample, through its denoising error, as in diffusion classifiers~\cite{li2023your,clark2023text}. For this analysis only, we enumerate the legal moves, draw each into the output image, measure how well the model denoises it under shared noise, and turn the errors into a policy with a softmax. Its most likely move is more often correct than any sampled readout of the same model, and it ranks the remaining moves in agreement with Stockfish (Appendix~\ref{app:implicit-policy}). In Figure~\ref{fig:implicit-policy}, every knight move gives check, yet most of the policy goes to Nh6+, the only one that wins.

\begin{figure*}[t]
    \centering
    \includegraphics[width=0.95\linewidth]{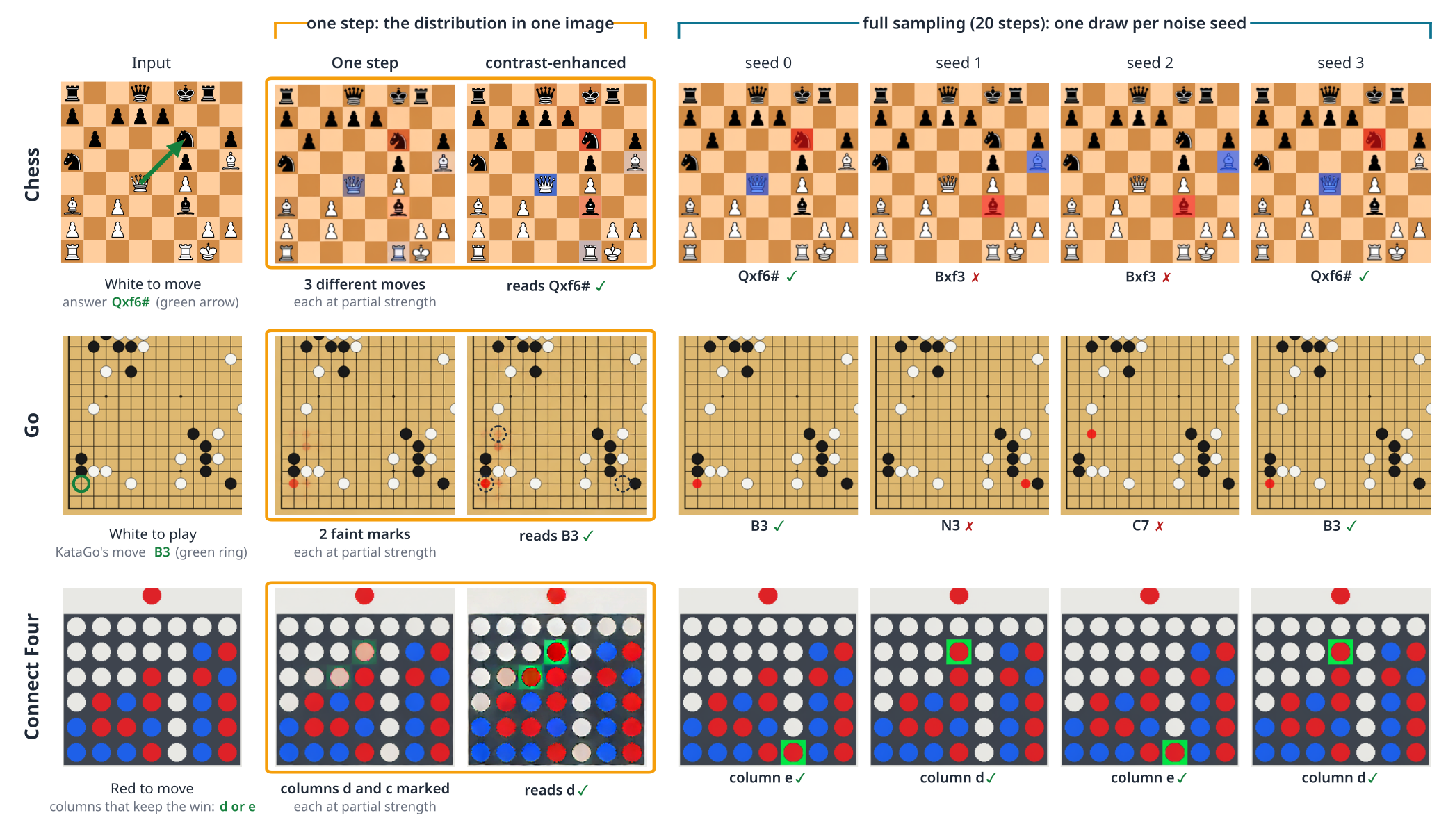}
    \caption{\textbf{One step reads the decision; full sampling draws one.} On held-out chess (top), Go (middle) and Connect Four puzzles (bottom), the one-step output marks several candidate moves with varying strengths, and its strongest mark is read as the decision. Full 20-step sampling instead commits to a single move, and different noise seeds commit to different ones. In the chess puzzle, half the samples play Bxf3 instead of the mate Qxf6\#, which the one-step output reads. Further quantitative results are in Appendix~\ref{app:readouts}.}
    \label{fig:fewstep}
\end{figure*}

\noindent\textbf{One step reads the decision.}
Training gives the model a distribution $\pi(m \mid c)$ over the moves $m$ of a board $c$, and the number of denoising steps only decides how this distribution is read out. The reason lies in the training objective: the best prediction of the clean output from a noisy one is the mean of all clean outputs that could have produced it, and at full noise, where the input carries no information about the move, this mean is the average of the move images $I_m$ weighted by their probabilities,
\begin{equation}
  \hat x_0 \;=\; \mathbb{E}[\,x_0 \mid c\,] \;=\; \textstyle\sum_m \pi(m \mid c)\, I_m
  \qquad \text{(full noise)}
  \label{eq:first-estimate}
\end{equation}
(Tweedie's formula~\cite{efron2011tweedie}; Proof in Appendix~\ref{app:why}). A single step therefore returns the whole distribution at once, and its strongest marks pick out the most likely move. This is why one step is the more accurate readout in every task that asks for a single decision (Figure~\ref{fig:fewstep}, Table~\ref{tab:chess_recall_vs_play}), and it also sets its limit: because the average hardly depends on the noise, repeated one-step outputs agree, and eight of them contain the answer for only 69.2\% of chess puzzles. Full sampling instead follows each noise draw to a single move, chosen in proportion to its probability. One sample is right less often, but eight of them contain the answer for 85.4\%. One step is thus the better answer and sampling the better source of candidates, and the gap between the two is headroom that no readout without the answer yet exploits. Averaging also explains why routes need full sampling: the average of several routes is not itself a route. In every task the decision is fixed within the first steps of denoising (Appendix~\ref{app:denoising}), as observed for video models solving mazes~\cite{newman2026video}; the remaining steps refine one candidate rather than search among several (Appendix~\ref{app:readouts}).

\noindent\textbf{Going beyond chess.}
The same formulation applies without modification to tasks with substantially different action spaces (Table~\ref{tab:extended_tasks}). What these tasks share is what the formulation needs: what the player observes can be shown in one image, and the action can be expressed as an edit of that image, for instance by marking the piece to move and its destination, the point to play, or the direction and strength of a shot. Within this class, the model must infer the properties of each task that are never stated. In pathfinding it is the cost of terrain: with costs hidden, the closed-loop image model finds the cheapest route on $87.5$\% of maps, against $59.2$\% for breadth-first search, which ignores terrain, while on uniform maps the video model, which plans several moves per clip, generalises better to longer paths. In Connect Four it is gravity: the disc lands in the correct row in all but one of 2{,}000 test positions for the image model and all but eight for the video model. In Go it is the rules of play: stones land on a legal point $99.8$\% of the time, and the model matches KataGo's move on $31.1$\% of positions, against $1.7$\% for copying the nearest training position. In billiards it is the physics of the table: the video model pots the ball on $47.9$\% of held-out layouts, against $19.0$\% for reusing the nearest training shot (Appendix~\ref{app:tasks}).

\noindent\textbf{Learning from the continuation.}
Drawing the future also helps the model decide the present. A video model trained to draw the whole solution line, replies included, predicts the next move more accurately than the same model trained on the first move alone, although only the first move is scored (24.6\% against 21.4\% in chess, with the same 400{,}000 puzzles and training budget), and the same holds on terrain for both models. Controlled experiments on a purpose-built pathfinding task show why (Appendix~\ref{app:horizon}): when the first move is decided by a single wall $k$ cells down one of two otherwise identical corridors, with equal training a model only trained on the first move ignores the wall beyond two cells, while one trained to draw the route uses it up to twelve cells away; the gain comes from training rather than from reading the drawn continuation at inference. Whether the drawn moves should also be played depends on the task. In deterministic tasks such as shortest paths, executing a whole clip is as accurate as predicting again after each move, with far fewer model calls. Against an opponent, a new prediction on the observed board is better than following the drawn continuation (52.5\% against 41.2\% for the next move on identical boards). Predicting the future can make each decision better, and against an opponent, each decision is best made on the board as it actually is.

%% file: sec/late-related.tex
\section{Related Work}

Having presented our formulation and results, we now place them in
the context of visual problem solving, sequential generation, and
learned decision-making. Our focus is on learning task rules and
action preferences from visual examples, and using this knowledge
to select the next best move (NBM).


Goren et al.~\cite{goren2025visual} show that visual diffusion models can be used to generate complete solutions to geometric problems in pixel space, including the inscribed square, Steiner tree, and maximum-area polygon. We extend visual solution generation to sequential decisions: in chess and pathfinding, successive predicted actions construct solutions, even when an action’s value depends on later consequences.

DiffThinker~\cite{he2025diffthinker} fine-tunes a pretrained image-editing
model to generate complete visual solutions to tasks including mazes,
TSP, Sudoku, and jigsaws. Video models have also shown signs of reasoning by generating maze trajectories, both zero-shot~\cite{wiedemer2025video} and after supervised fine-tuning~\cite{yang2025vrbench}. These works establish visual generation as a means of producing complete solutions or their unfolding trajectories. Our single-move model makes a different
commitment: it selects an action without generating its future continuation. Our horizon model connects these formulations by generating short continuations, allowing us to compare decisions made with and without an explicit visual trajectory.

Wewer et al.~\cite{wewer2025spatial} are closely related where they show that sequential generation can improve reasoning in denoising models. Their Spatial Reasoning Models use uncertainty to order the generation of spatial variables in visual Sudoku and other constraint tasks. Their approach selects which unknown part of a solution to resolve next, prioritizing
variables that can be predicted with greater confidence. Our experiments
address action selection according to its contribution to the eventual
outcome. The sacrifice in chess and the longer but cheaper terrain
route illustrate choices that cannot be judged by immediate gains
alone.

Ezquerro and Özdenizci~\cite{ezquerro2026relational} improve logical
constraint satisfaction through object-centric relational abstractions that guide the generation of complete solution images.
They demonstrate improvements on tasks including colored Sudoku,
Tangram, and logic-based face generation. Their contribution centers
on adding representations of objects and their relationships.
Our models acquire their decision-making capabilities through
fine-tuning pretrained visual diffusion models, without additional
object-discovery or relational modules.

Learning strong actions without inference-time search has also been
studied in chess using symbolic representations. Ruoss et
al.~\cite{ruoss2024amortized} train on engine-annotated positions to reach grandmaster-level play without inference-time search. DiffuSearch~\cite{ye2025implicit} uses discrete diffusion to predict future states and actions alongside the next move, using the generated continuation as a form of implicit search. These methods use chess-specific encodings and orders of magnitude more engine-labelled positions. We ask whether a general visual generator can learn valid and effective actions from pixels across tasks ranging from chess to billiards. Unlike a constrained chess output vocabulary, our pixel readout must learn legal and effective movements from examples alone. Our horizon variant further connects our approach to DiffuSearch by expressing predicted actions and future states visually.


Our horizon-based execution also relates to diffusion models for
planning and control. Diffuser~\cite{janner2022planning} generates
trajectories through denoising using receding-horizon planning, while
Diffusion Policy~\cite{chi2025diffusion} predicts action sequences
and executes part of each sequence before predicting again. Our
horizon model follows the same execution principle: execute the
first predicted action, observe the resulting state, and predict
again. Here, states and action markings share a visual representation,
and actions are decoded from generated pixels. In adversarial tasks,
the generated continuation also includes predicted opponent replies.
This connection places our comparison of single-move and horizon-based
prediction within the broader study of learning to act through
generative models.

%% file: sec/conclusion.tex
\section{Discussion and Conclusions}

In this work, we show that visual diffusion models can learn to decide what to do next by learning to draw the next move. This simple formulation supports a surprisingly wide range of decisions: sacrificing a rook for a payoff several moves later, taking a longer route to avoid costly terrain, or selecting a shot in billiards. In sequential tasks, the model builds a solution one decision at a time, without explicitly searching through future continuations.

A natural question is why we should generate a move rather than classify it. In chess, a classifier with an output designed to score pairs of board squares can outperform our model. The advantage of generation is that the same output representation works across tasks: the model draws a chess move, a route, or a billiard shot directly on the image. This goes beyond reusing an architecture. When we train a single model on Connect Four and pathfinding together, with the same prompt for both, it matches the single-task model on Connect Four and stays within a few points of it on pathfinding. The appendix reports these comparisons in detail.

The most intriguing finding is that the first denoising step can be more useful for choosing a move than a fully denoised sample. It shows several possible moves as overlapping marks. In our experiments, reading the strongest marks gives a more accurate single decision than drawing one fully denoised sample, without any additional training. Full denoising remains useful when we want diverse candidates. For a model that learns to act by drawing, the first sketch can already be the better guide to what to do next.

%% file: sec/X_suppl.tex
\clearpage
\setcounter{page}{1}
\maketitlesupplementary
\appendix

\section{Tasks and data}
\label{app:tasks}

Each task presents a rendered state and asks the model to draw an action. The video model also draws subsequent actions and states. Prompts never state the rules, costs, or physics. Except for chess and Go, the prompt is simply ``Continue.'' An exact solver or strong engine supplies the labels used to build the data and score predictions. Below, we specify each task, its rendering, data and splits, and evaluation.

\subsection{Chess}
\label{app:task-chess}

\noindent\textbf{Problem.} Positions come from the Lichess puzzle database. Each puzzle starts after the opponent's last move and has an engine-verified solution line in which only the first move of the side to move is forced to be unique. The target is that first move.

\noindent\textbf{Rendering.} Boards are drawn from White's side at $256{\times}256$ pixels, without coordinates. The model marks a move by tinting the origin square blue and the destination square red. The image model's prompt is ``\emph{White to move. Highlight the best move: the origin square in blue and the destination square in red.}'' (Black analogously). The video model additionally draws green dots on the legal destination squares of the moving piece and then shows the board after the move, as described in \cref{app:models}.

\noindent\textbf{Data and splits.} Puzzles are assigned to splits by a hash of the game they come from, so all puzzles of one game share a split. Any puzzle whose start position appears in validation or test (same pieces, side to move, castling and en passant rights) is removed from training, which also excludes transpositions. Validation and test keep only well-established puzzles (rating deviation at most 90 and at least 200 plays) and are stratified over eight rating bands, from below 1000 to above 2500. This gives 5{,}808{,}734 training, 2{,}000 validation and 10{,}000 test puzzles. The models of \cref{tab:chess_recall_vs_play} train on the first three million training puzzles. The variants and analyses in this supplement use the first million unless stated otherwise. For the video model, each puzzle becomes a line of up to eight solution plies (mean 3.6).

\noindent\textbf{Engine labels.} For the implicit-policy analysis and for contrastive guidance we score every legal move with Stockfish~19 (all moves at depth 12, at most 4M nodes per position, one thread). Scores are converted to a win probability for the side to move with the Lichess formula, and forced mates rank above every centipawn score. The engine's best move equals the puzzle answer on 99.8\% of test positions.

\noindent\textbf{Metrics and baselines.} A prediction is correct when the move read from the generated image (\cref{app:eval}) equals the puzzle answer. Promotion pieces are ignored. We report two subsets: \emph{sacrifices} (1{,}001 test puzzles tagged as such by Lichess) and \emph{far from training}, the test puzzles whose nearest training position differs on 13 to 16 of the 64 squares (1{,}757 puzzles against the three million training puzzles, 2{,}403 against the first million). Retrieval baselines and their construction are in \cref{app:recall}.

\begin{figure*}[!t]
\centering
\includegraphics[width=\linewidth]{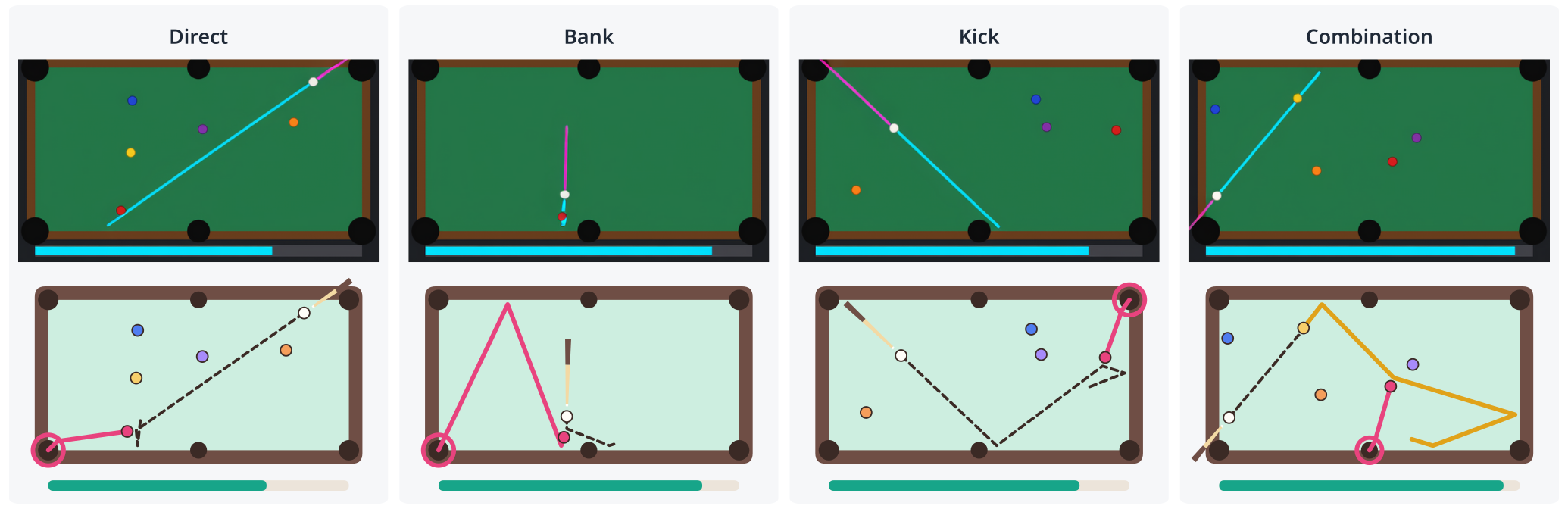}
\caption{\textbf{Billiards decisions and outcomes.} Each column shows a held-out layout for one shot type: direct, bank, kick, or combination. \emph{Top:} The video model's unaltered output marks the shot with an aim line from the cue ball, a cue stick behind it, and a power bar whose fill indicates shot strength. \emph{Bottom:} The corresponding simulator rollout. The dashed line traces the cue ball, and the ring marks the pocket that receives the red ball. The bank, kick, and combination layouts are hard, where no direct pot exists.}
\end{figure*}

\subsection{Shortest paths and terrain}
\label{app:task-path}

\noindent\textbf{Problem.} An agent on an $N{\times}N$ grid ($N\le12$) must reach a goal. In the uniform variant every step costs the same and the target is a shortest path. In the terrain variant, cells are floor, mud or sand, entering them costs 1, 2 or 4, and the target is a cheapest path. Walls cannot be entered. The costs are never shown and must be inferred from examples.

\noindent\textbf{Rendering.} Maps are drawn at $384{\times}384$ pixels with 32-pixel cells, so a cell is exactly one video-model token and $2{\times}2$ image-model tokens. The agent is a blue disc, the goal a red ring, and visited cells are green. Mud has stripes and sand has dots, so the two can be told apart without colour alone. A rule-free decoder reads every cell by matching its centre, ring and corner colours to the known palette.

\noindent\textbf{Formats.} The image model operates in a closed loop: it marks the next cell with a red disc, the move is applied, and the resulting state becomes the next input. The video model generates segments of up to eight moves, each with a decision frame and the resulting state. Longer routes chain these segments. The receding-horizon variant executes only the first decision from each segment (\cref{app:eval}).

\noindent\textbf{Data.} Maps come from three families: random blocks (15 to 32\% wall density), walls with one or two gaps, and rooms made by recursive division. Terrain maps cover 25 to 45\% of free cells with mud or sand. Targets are computed by breadth-first search (Dijkstra for terrain), breaking ties by fewest turns, and evaluation accepts any optimal path. Each variant has 200{,}000 training maps and 600 validation maps. The uniform variant has 1{,}000 in-distribution test maps and three out-of-distribution tests of 500 maps each: larger maps ($N$ = 11 or 12), longer paths (21 to 32 steps) and mazes, a family never seen in training. The terrain variant has 1{,}000 test maps and a probe set of 500 maps on which every cheapest path either crosses a band of costly terrain (250 maps) or goes around it (250 maps). No evaluation map appears in training, and the evaluation sets are pairwise disjoint.

\noindent\textbf{Metrics and baselines.} A route is valid if it is connected, avoids walls and revisits, and changes nothing off the path. It is optimal if it is valid and has the optimal length (terrain: the optimal cost). On the probe set, a route is correct if it is valid and crosses the band exactly when the cheapest paths do. The terrain-blind baseline follows a shortest path in steps. It is cost-optimal on 59.2\% of terrain test maps and, by construction of the probe set, correct on half of the probe maps. The pretrained models without fine-tuning never produce a valid route. On the maze test, the image and video models are optimal on 78.4 and 84.4\% of maps.

\subsection{Connect Four}
\label{app:task-c4}

\noindent\textbf{Problem.} Standard $7{\times}6$ Connect Four: the model chooses a column for the side to move. That a disc falls to the lowest free row is never stated.

\noindent\textbf{Rendering.} Boards are $256{\times}256$ pixels with 32-pixel cells. Red moves first and blue second, and a disc in the top margin shows whose turn it is. A move is drawn as the mover's disc in its landing cell, tinted green.

\noindent\textbf{Data.} Positions come from self-play games in which each player plays a random legal move with a probability drawn from $\{0, 0.05, 0.1, 0.2, 0.35, 0.5\}$ and otherwise a best or value-preserving move. We keep only critical positions, in which not every column has the same game-theoretic value. Labels come from Pascal Pons' exact solver. The label is a best column, and evaluation accepts any best column. There are 200{,}000 training, 1{,}000 validation and 2{,}000 test positions, plus a hard test set of 1{,}000 positions with neither an immediate win nor an immediate threat. Positions are deduplicated up to mirror symmetry within and across splits, and each evaluation set takes at most one position per game.

\noindent\textbf{Metrics and baselines.} \emph{Legal, correct landing row}: the marked column is not full and the disc is drawn in the row where it would land. \emph{Keeps the win}: on the hard test set, the chosen move preserves the game-theoretic value of the position, so a won position stays won and a drawn one stays drawn. Of the 1{,}000 hard positions, 936 are won. The one-move heuristic wins immediately if it can, otherwise blocks the opponent's immediate win, otherwise plays the most central legal column.

\subsection{Go}
\label{app:task-go}

\noindent\textbf{Problem.} Full $19{\times}19$ Go. The model marks the move KataGo prefers.

\noindent\textbf{Data and labels.} We take games between strong amateurs (7 dan, or both players at least 6 dan) from the public KGS archive, keeping even games of at least 80 moves played up to 2015 without bots. Positions are sampled from the middle of each game. Each position is searched by KataGo (network \texttt{b11c768}, 400 visits, Chinese rules, komi 7.5), and the label is the move with the most visits. Positions whose outcome is already clear are filtered out. The smaller training set has 100{,}000 positions from 44{,}191 games, and the larger one has 1{,}000{,}000 positions from 91{,}253 games. Both share 1{,}000 validation and 2{,}000 test positions, and no game or board position is shared between splits.

\noindent\textbf{Rendering and prompt.} Boards are $384{\times}384$ pixels, and the move is a red disc on the chosen point. The prompt is ``\emph{Black to play. Mark the best move with a red dot.}'' (White analogously).

\noindent\textbf{Metrics and baselines.} A prediction is correct when the decoded point equals KataGo's move. With about 296 legal points per position, a random legal move is correct 0.35\% of the time in expectation. The single random draw reported in \cref{tab:extended_tasks} is correct on 15 of the 2{,}000 positions (0.7\%). Copying the move of the nearest training position is correct 1.7\% of the time. For reference, KataGo's raw policy network alone (no search) matches its own searched move 74.0\% of the time, and the move played in the game matches 39.5\%.

\subsection{Billiards}
\label{app:task-billiards}

\noindent\textbf{Problem.} A table holds the cue ball, a red target ball and up to four other balls. The model chooses a shot, a cue direction and a power, and succeeds if the red ball is potted and the cue ball is not.

\noindent\textbf{Simulator and solver.} Shots are simulated by a deterministic 2D simulator without spin: rolling friction, elastic ball collisions (restitution 0.95), cushions (restitution 0.8) and six pockets. For each layout we simulate 12{,}960 shots (every $0.25^\circ$ and nine powers) and take as label the centre of the widest successful range of angles that is robust to small changes in power. Shots include direct pots, banks, kicks, kisses and combinations.

\noindent\textbf{Rendering.} Tables are $512{\times}288$ pixels. A shot is drawn as a cyan aiming line from the cue ball to where its centre would first meet a cushion, drawn under the balls, together with a magenta cue stick behind the cue ball and a power bar under the table. A rule-free decoder reads the angle from the line and stick and the power from the filled part of the bar. On clean renders its angle error has a median of $0.003^\circ$. In the video model's clip, the decision frame is followed by the simulated motion of the balls.

\noindent\textbf{Data and metrics.} There are 200{,}000 training, 1{,}000 validation and 2{,}000 test layouts, and a mirrored copy of every test layout. No layout is shared between splits up to the table's symmetries. A shot counts as \emph{potted} if the decoded shot succeeds in the simulator. The nearest-shot baseline finds the training layout whose cue and red balls are closest (including mirrored layouts) and reuses its shot. It pots 19.0\% of test layouts, and a random shot 0.7\%. On the mirrored layouts the one-step image and video models pot 28.2 and 49.1\%, against 28.0 and 47.9\% on the originals.

\subsection{N-queens}
\label{app:task-nqueens}

\noindent\textbf{Problem and rendering.} On a partly filled $N$-queens board, the model places the next queen so that the board can still be completed with no two queens sharing a row, column or diagonal. Boards with $N\le12$ are drawn on a fixed $12{\times}12$ grid. Qwen-Image-2.1, fine-tuned with rank-64 LoRA, receives the prompt ``Continue.'' and marks the next queen in red. The action is read from the strongest red mark, without a minimum-strength threshold.

\noindent\textbf{Data and evaluation.} We train on 200{,}000 boards for 24k steps, using sizes $N=8,9,10$. Whole solution families related by rotation or reflection are held out from training for validation and test: 4 of 12 families at $N=8$, 16 of 46 at $N=9$, and 30 of 92 at $N=10$. Sizes 11 and 12 are never seen in training. No test board or any of its rotations or reflections occurs in training. Exact enumeration supplies the labels: any placement that still permits a complete solution is correct. We select one sampling step on validation.

\noindent\textbf{Results.} On held-out solution families, the model trained on the first move only makes a completable move on 85.2\% of boards and a legal placement on 99.4\%. A random non-attacked cell is completable on 71.2\% and copying a move from the nearest training board on 62.3\%. The copier falls from 96.8\% on training families to 62.3\% on held-out families, while the model's advantage over the random non-attacked choice remains similar (+15.5 and +14.0 points). On the never-trained sizes 11 and 12, the model reaches 79.2\% against 72.6\% for that random baseline, with 97.0\% legal placements. On hard boards, it reaches 62.0\% against 49.1\%. At 20 sampling steps, held-out-family accuracy is 78.8\%, below the validation-selected one-step result. The controlled horizon test is in \cref{app:horizon}.

\section{Models and training}
\label{app:models}

\noindent\textbf{Image model.} We fine-tune Qwen-Image-2.1 with LoRA (rank 64, $\alpha=64$) on the attention projections and image MLP layers of its transformer. The input image is encoded as the condition and the model is trained with the rectified-flow velocity loss to produce the edited image.

\noindent\textbf{Video model.} We fine-tune Wan2.2-TI2V-5B with LoRA of the same rank on its attention and feed-forward layers. The input state occupies the first frame and is excluded from the loss. Each clip contains 17 blocks of four frames (65 frames): block 0 shows the input, and for each of up to eight moves, a decision block shows the marked move and the next block shows its resulting state. The final state is held after the last move. At 256 pixels, one latent token covers one chess square. In chess, each decision block additionally marks the legal destination squares of the moving piece with green dots. The decoder reads the move from the blue and red squares only (\cref{app:eval}). The dots are used to test whether the drawn destinations match the piece's legal moves, and the image models are trained without them.

\begin{table}[htb]
\centering\footnotesize
\begin{tabular*}{\linewidth}{@{\extracolsep{\fill}}llrrc@{}}
\toprule
Task & Model & Steps & Batch & Resolution \\
\midrule
Chess, 3M & image & 375k & 8 & 256 \\
Chess, 3M & video & 187.5k & 16 & 256 \\
Chess, 1M & image & 125k & 8 & 256 \\
Chess, 1M & video & 62.5k & 16 & 256 \\
Terrain path & image & 24k & 8 & 384 \\
Paths & video & 5k & 16 & 384 \\
Connect Four & image & 12k & 16 & 256 \\
Connect Four & video & 10k & 16 & 256 \\
Go, 100k & image & 20k & 8 & 384 \\
Go, 1M & image & 125k & 8 & 384 \\
Billiards & image & 20k & 16 & $512{\times}288$ \\
Billiards & video & 12k & 16 & $512{\times}288$ \\
\bottomrule
\end{tabular*}
\caption{Training configurations for the models used in the main results.}
\label{tab:supp-runs}
\end{table}

\noindent\textbf{Training.} All runs use AdamW with learning rate $10^{-4}$, cosine decay, 200 warm-up steps, weight decay $10^{-4}$ and gradient clipping at 1. \Cref{tab:supp-runs} lists the runs behind the main results. The image models in the main tables see only the best-move prompt. The models used for contrastive guidance (\cref{app:guidance}), chess at one million puzzles and Go at 100{,}000 positions, are instead trained on a mixture of prompts: half of the examples ask for the best move, and the rest for any legal move or a plausible but losing move.

\section{Inference and evaluation}
\label{app:eval}

\noindent\textbf{Sampling.} The image model is sampled with a 20-step Euler flow-matching sampler, the video model with UniPC for 30 steps (flow shift 3, no classifier-free guidance).

\noindent\textbf{Reading the action.} We decode actions from pixels without using task rules. In chess, we fit the blue and red tint added to each square and select the strongest blue square as the origin and the strongest red square as the destination. The other tasks use the same principle with their respective palettes (\cref{app:tasks}). The strongest mark of each kind determines the action regardless of its strength. We do not constrain the readout to legal actions, so an illegal action counts as a failure. For video, we average each block's two central frames and check each decision against the board the model imagined immediately before it.

\noindent\textbf{Closed loop and receding horizon.} To play a full line or route, the image model is called once per move: its move is applied and the true new state is rendered for the next call. The video model generates up to eight moves at a time. In the receding-horizon variant, only the first decision of each clip is executed and a new clip is generated from the resulting state.

\noindent\textbf{Contrastive guidance.} Guided sampling is described and evaluated in \cref{app:guidance}. With $w=0$ the sampler is the unguided one.

\noindent\textbf{Majority voting.} We draw 8 samples with different noise in one batch and keep the most frequent decoded move, breaking ties by sample order. For scoring, samples whose drawing is not a legal move are pooled into a single ``no move'' vote, which is never correct, so a prediction fails whenever this vote wins. Legality is thus checked by the evaluation only, and pooling is the stricter convention compared with the mode over raw decoded moves of \cref{eq:vote}. In billiards, the vote is the circular median of the decoded angles and the median power. On the transformed and changed boards of \cref{app:recall} we use 4 samples.

\noindent\textbf{Few-step sampling.} We also sample with fewer denoising steps. For \cref{tab:chess_recall_vs_play} and \cref{tab:supp-recall-1m}, the number of steps is selected on the validation set only: among 1, 2, 3 and 4 steps and the default, the most accurate setting, with ties going to fewer steps. This selects one step for both models of \cref{tab:chess_recall_vs_play}, and two steps for the image and one for the video model at one million puzzles. In \cref{tab:extended_tasks}, every row uses full sampling, and the ``+ one-step'' rows of the single-decision tasks (Connect Four, Go, billiards) use exactly one step, without selection, and \cref{tab:supp-steps-tasks} reports every step count. With contrastive guidance, fewer steps do not help (\cref{tab:supp-steps}), so we report no few-step setting with guidance.

\section{Train with diffusion, decide in one step}
\label{app:readouts}

Few-step sampling, majority voting and the implicit policy all beat a single full sample (\cref{tab:chess_recall_vs_play}). This section shows that there are three ways of reading one object: the distribution over actions that the model learns in training. The model's estimate of the clean output at full noise is the average of the action images, each weighted by its probability, so its strongest marks give the most likely action. One denoising step returns this average. Many steps instead turn it into a single draw from the distribution, which is right less often. We derive this from the training objective (\cref{app:why}), measure the distribution directly (\cref{app:implicit-policy}), and test the consequences in chess (\cref{app:evidence}) and in the other tasks (\cref{app:beyond}). \Cref{app:implications} discusses what follows for inference.

\begin{figure}[tb]
\includegraphics[width=\linewidth]{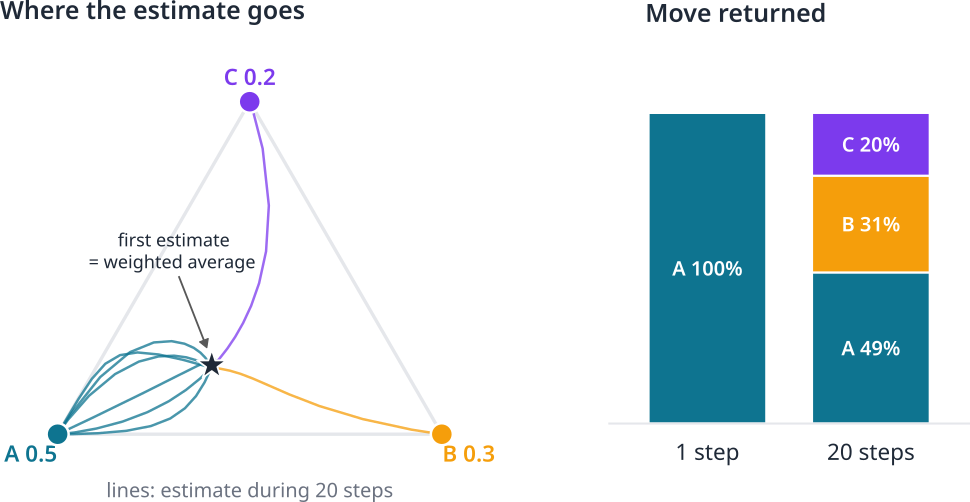}
\caption{\textbf{One step reads the decision, many steps sample it.} An exact toy model with three moves (probabilities 0.5, 0.3 and 0.2) whose target images each tint one square, so that the moves are the corners of a triangle. \emph{Left:} for every noise draw, the first estimate of the clean output is the probability-weighted average (star). During 20 steps, the estimate travels from there to one corner (lines). \emph{Right:} the move returned over 2{,}000 noise draws.}
\label{fig:supp-toy}
\end{figure}

\Cref{fig:supp-toy} shows the idea on an exact toy model with three moves of probability 0.5, 0.3 and 0.2. From every noise draw, the first estimate is the same point, the weighted average of the three move images, and its strongest mark is the most likely move. Twenty steps instead carry each noise draw to one of the three moves, in proportion to their probabilities. If the answer follows the model's probabilities, the first estimate is right half of the time and a 20-step sample $0.5^2 + 0.3^2 + 0.2^2 = 0.38$ of the time, since both the ground truth and the model must independently choose the same answer.

\subsection{Why one step reads the most likely action}
\label{app:why}

\noindent\textbf{Setup.} For a board $c$ and a noise level $\sigma \in (0, 1]$, training corrupts the clean output $x_0$ into $x_\sigma = (1-\sigma)\,x_0 + \sigma\epsilon$ with $\epsilon \sim \mathcal N(0, I)$, and the network $v$ predicts the velocity $\epsilon - x_0$. Its estimate of the clean output is $\hat x_0 = x_\sigma - \sigma v$. We write $I_m$ for the target image of action $m$ and $m^\star$ for the correct action, and treat the targets of a board as a mixture: given $c$, the clean output is $I_M$ with $M \sim \pi(\cdot \mid c)$, where $\pi_m = \pi(m \mid c)$ is the probability the model assigns to action $m$. This ignores variation within the image of one action.

\noindent\textbf{Lemma 1 (the model predicts a conditional mean).} The training loss $\mathbb{E}\,\|v(x_\sigma, c) - (\epsilon - x_0)\|^2$ is minimised by $v^\star = \mathbb{E}[\epsilon - x_0 \mid x_\sigma, c]$, and then
\begin{equation}
  \hat x_0(x_\sigma, c) = \mathbb{E}[\,x_0 \mid x_\sigma, c\,].
  \label{eq:tweedie}
\end{equation}
\emph{Proof.} Write $X = (x_\sigma, c)$ for the input, $Y = \epsilon - x_0$ for the target and $g(X) = \mathbb{E}[Y \mid X]$. For any predictor $v$,
\begin{align*}
  \mathbb{E}\|v(X) - Y\|^2
  &= \mathbb{E}\|v(X) - g(X)\|^2 + \mathbb{E}\|g(X) - Y\|^2 \\
  &\quad + 2\,\mathbb{E}\bigl[(v(X) - g(X))^\top (g(X) - Y)\bigr].
\end{align*}
The last term is zero: conditioned on $X$, the factor $v(X) - g(X)$ is fixed and $g(X) - Y$ has mean zero. The second term does not depend on $v$, so the loss is smallest when $v = g$. Since $x_\sigma = (1-\sigma)\,x_0 + \sigma\epsilon$, the target can be written as $Y = (x_\sigma - x_0)/\sigma$, so
\begin{align*}
  v^\star &= \bigl(x_\sigma - \mathbb{E}[x_0 \mid X]\bigr)/\sigma, \\
  \hat x_0 &= x_\sigma - \sigma v^\star = \mathbb{E}[x_0 \mid X]. \qquad\square
\end{align*}
This is Tweedie's formula~\cite{efron2011tweedie} written for flow matching.

\noindent\textbf{Proposition 1 (the first estimate is a weighted average of actions).} At $\sigma = 1$, for every noise draw,
\begin{equation}
  \hat x_0(x_1, c) = \sum_m \pi_m\, I_m .
  \label{eq:superposition}
\end{equation}
\emph{Proof.} At $\sigma = 1$ the input $x_1 = \epsilon$ is pure noise, independent of $x_0$ given $c$, so by \cref{eq:tweedie}
\begin{equation*}
  \hat x_0(x_1, c) = \mathbb{E}[x_0 \mid c] = \sum_m \pi_m I_m . \qquad\square
\end{equation*}
A one-step sampler outputs exactly this estimate, $x_1 - v(x_1, c) = \hat x_0(x_1, c)$. The image model's two-step schedule ($\sigma = 1 \to 0.02 \to 0$) returns nearly the same image, since its second step starts almost at $\sigma = 0$.

\noindent\textbf{Reading the average.} Suppose the marks add up, $I_m = B + \beta\, e_{f(m)} + \rho\, e_{t(m)}$, where $B$ is the board, $e_s$ the tint of square $s$, and $f(m)$ and $t(m)$ the origin and destination of move $m$. By \cref{eq:superposition}, square $s$ is then blue with strength $\pi_{\mathrm{from}}(s) = \sum_{m:\,f(m) = s} \pi_m$ and red with strength $\pi_{\mathrm{to}}(s) = \sum_{m:\,t(m) = s} \pi_m$, and reading the strongest blue and the strongest red square returns the most likely move whenever that move holds more than half of the probability. Below that, the readout can pair the origin of one move with the destination of another when several moves share a square. In practice this is rare (\cref{app:evidence}).

\noindent\textbf{Reading the most likely action beats sampling.} A sampler that is integrated to the end draws an action from $\pi$, which is correct with probability $\pi_{m^\star}$. Reading the most likely action is correct if $\arg\max_m \pi_m = m^\star$. If the model is calibrated, i.e.\ among the boards to which it assigns probabilities $\pi$ the correct action is $m$ on a fraction $\pi_m$ of them, the two accuracies over boards are $\mathbb{E}[\sum_m \pi_m^2]$ and $\mathbb{E}[\max_m \pi_m]$, and since every $\pi_m \le \max_{m'} \pi_{m'}$,
\begin{equation}
  \sum_m \pi_m^2 \;\le\; \max_{m'} \pi_{m'} \sum_m \pi_m \;=\; \max_m \pi_m ,
  \label{eq:sample-vs-mode}
\end{equation}
with equality only when $\pi$ is uniform over the actions it allows. The gap vanishes for a certain and for a uniform model and is largest in between. A majority vote over samples converges to the most likely action, slowly when the leading actions are close, and at least one of $n$ samples is correct with probability $\mathbb{E}[1 - (1 - \pi_{m^\star})^n]$, which approaches one even when $m^\star$ is not the most likely action.

\subsection{Measuring the distribution: the implicit policy}
\label{app:implicit-policy}

To test these predictions we need $\pi$ itself. A conditional diffusion model can score outputs it did not sample, through how well it denoises them. Diffusion classifiers use this to pick the class whose prompt makes an image easiest to denoise~\cite{li2023your,clark2023text}. We score every legal move in this way. Unlike the readout of \cref{app:eval}, this analysis is given the legal moves.

\noindent\textbf{Scoring a move.} We encode $I_m$ with the VAE into the latent $x_0^{(m)}$, add noise, $x_\sigma^{(m)} = (1-\sigma)\, x_0^{(m)} + \sigma\, \epsilon_k$, and average the error of the predicted velocity,
\begin{equation*}
E(m) = \frac{1}{|\Sigma|\,K} \sum_{\sigma \in \Sigma} \sum_{k=1}^{K}
\frac{1}{d}\, \bigl\| v\bigl(x_\sigma^{(m)}, c\bigr) - \bigl(\epsilon_k - x_0^{(m)}\bigr) \bigr\|^2 ,
\end{equation*}
over $\Sigma = \{0.8, 0.9\}$ and $K = 8$ noise draws, where $d$ is the size of the latent. All moves of a board are scored with the same draws $\epsilon_k$, so differences in $E(m)$ come only from the moves. The implicit policy is $q(m) \propto \exp(-E(m)/T)$ over the legal moves.

\noindent\textbf{The error reads $\pi$.} Fix a move $m$ and one noise draw $\epsilon$, and write $z = (1-\sigma)\, I_m + \sigma\epsilon$ for the noisy version of the move's image. The velocity the model is trained to predict is $\epsilon - I_m = (z - I_m)/\sigma$, and the velocity it predicts can be written as $v = (z - \hat x_0)/\sigma$, where $\hat x_0$ is its estimate of the clean output. Subtracting the two, the error of the move becomes
\begin{equation}
  \bigl\| v - (\epsilon - I_m) \bigr\|^2 \;=\; \frac{1}{\sigma^2}\, \bigl\| I_m - \hat x_0 \bigr\|^2 ,
  \label{eq:ip-identity}
\end{equation}
for any network: the error of a move is the distance between the move and the model's estimate.

For the optimal model, \cref{eq:tweedie} makes this estimate a weighted average of all moves,
\begin{equation*}
  \hat x_0 = \sum_k w_k I_k ,
  \qquad
  w_k \;\propto\; \pi_k\, \exp\!\Bigl(-\frac{\|z - (1-\sigma) I_k\|^2}{2\sigma^2}\Bigr),
\end{equation*}
where $w_k$ is the probability of move $k$ given the noisy input, by Bayes' rule: the prior $\pi_k$ times how well move $k$ explains $z$.

With two candidate moves $m$ and $k$ at distance $d = \|I_m - I_k\|$, the error of $m$ is $(1 - w_m)^2 d^2 / \sigma^2$, and
\begin{align}
  \log\frac{w_m}{w_k}
  &= \underbrace{\log\frac{\pi_m}{\pi_k}}_{\text{belief}}
   + \underbrace{\frac{(1-\sigma)^2}{2\sigma^2}\, d^2
   + \frac{1-\sigma}{\sigma}\,\langle \epsilon,\, I_m - I_k \rangle}_{\text{evidence in } z} .
  \label{eq:ip-logodds}
\end{align}
For a fixed noisy input, a larger $\pi_m$ gives a larger $w_m$ and therefore a smaller error. Each move is scored on its own noisy input, so between two moves the last term enters with opposite signs, and the comparison holds on average over the noise rather than for every draw. This is why we average over $K$ draws, and the move with the lowest average error is then the one the model believes in most. With more than two candidates, moves that share a square lie closer to the average than isolated ones, so the errors follow $\pi$ only approximately, and \cref{app:evidence} measures how closely. The noise level sets the balance between the two parts: near $\sigma = 0$ the evidence dominates and every move is denoised perfectly, while at $\sigma = 1$ the evidence vanishes and the error is the distance to the average of \cref{eq:superposition}. Intermediate levels such as ours keep both.

\noindent\textbf{Results.} On the first 1{,}000 test puzzles, the move with the lowest error is the answer for 67.0\% of the puzzles and among the three lowest for 88.1\% (a random legal move: 5.6\% and 16.1\%, the pretrained model scored with four noise draws: 6.7\% and 15.5\%). Within each puzzle, the errors also rank the other moves in agreement with Stockfish, among quiet moves as well and after removing the effects of piece type, captures and checks.

\subsection{Evidence in chess}
\label{app:evidence}

We test the consequences on the image model of \cref{tab:chess_recall_vs_play}, on the first 1{,}000 test puzzles, for which we have its one-step outputs, eight 20-step samples and the implicit policy. To compare the policy with the samples, we fit its temperature to the moves the samples choose, without using the answers ($T = 0.0017$).

\begin{figure}[htb]
\centering
\includegraphics[width=\linewidth]{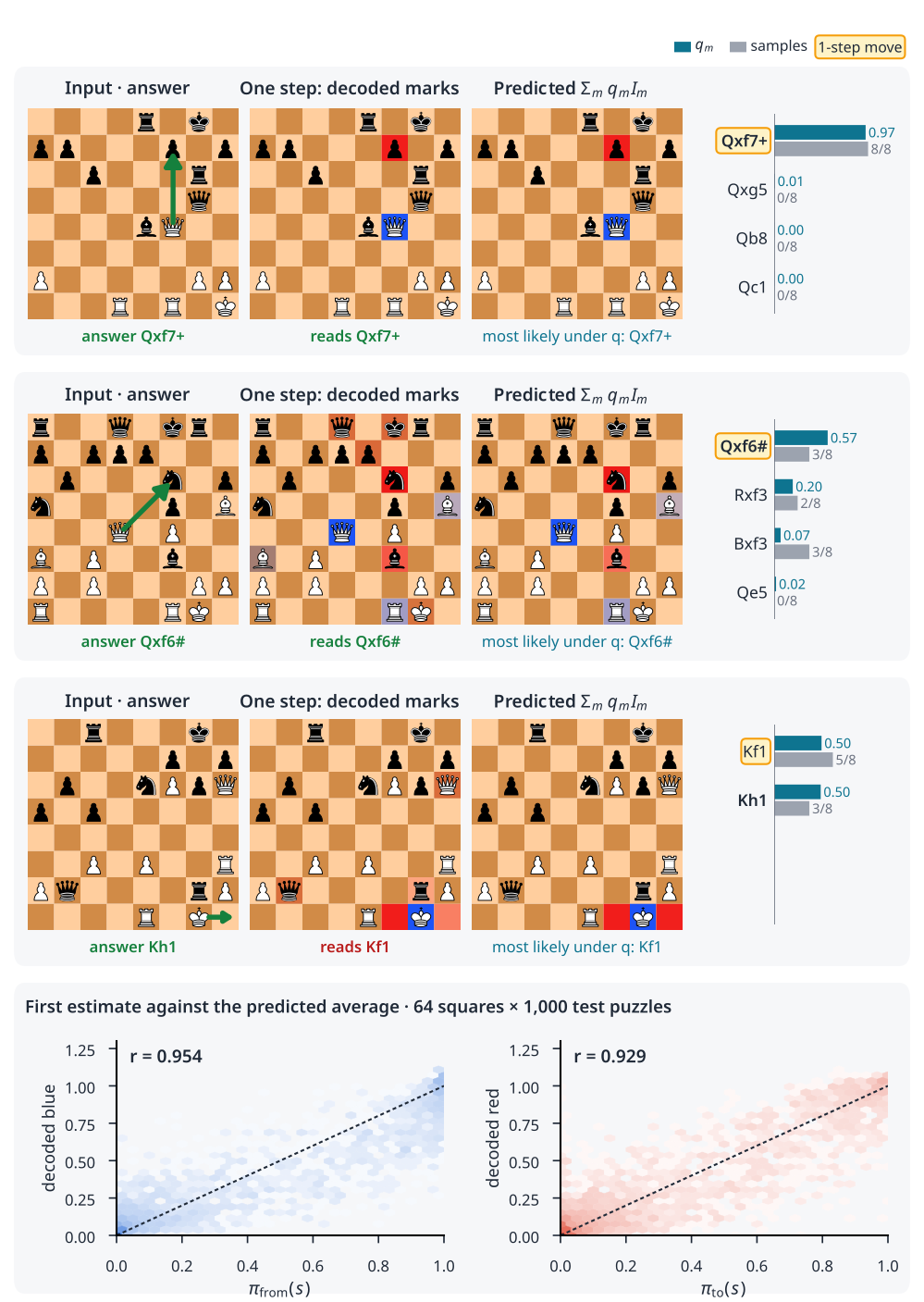}
\caption{\textbf{One step draws the whole distribution.} \emph{Top:} for each puzzle, the marks of the model's one-step output closely match the average of all moves weighted by the implicit policy $q$ (Proposition~1): each candidate appears with its probability, and reading the strongest mark gives the most likely move, which a single 20-step sample often misses (bars: $q$ vs.\ the share of 8 samples). \emph{Bottom:} the same comparison for every square of 1{,}000 puzzles: the blue (red) tint of each square in the one-step output against $\pi_{\mathrm{from}}$ ($\pi_{\mathrm{to}}$), the probability that a move starts (ends) there. The dashed line is an exact match.}
\label{fig:supp-superposition}
\end{figure}

\noindent\textbf{The first estimate is the predicted average.} \Cref{fig:supp-superposition} shows one-step outputs next to the average $\sum_m q_m I_m$ that Proposition~1 predicts from the implicit policy. Across all 64 squares of the 1{,}000 puzzles, the blue and red strengths of the one-step outputs follow the predicted strengths $\pi_{\mathrm{from}}$ and $\pi_{\mathrm{to}}$ closely (Pearson correlation 0.95 and 0.93). As Proposition~1 predicts, the output barely depends on the noise: two noise draws give the same move on 89.4\% of the puzzles. The move read from the output is the most likely move under $q$ on 80.2\% of the puzzles. The mispairing described in \cref{app:why} is predicted on only 40 of the 1{,}000 puzzles, and on these the output shows the mispaired move only 5 times and the most likely move 17 times: the network keeps origins and destinations together better than additive marks would.

\begin{table}[htb]
\centering\footnotesize
\setlength{\tabcolsep}{4pt}
\begin{tabular*}{\linewidth}{@{\extracolsep{\fill}}lcc@{}}
\toprule
Readout & Measured & Expected \\
\midrule
20-step sample & 57.1 & 55.6 \\
Vote over 8 samples & 63.8 & 64.7 \\
Two-step sample & 63.3 & -- \\
Implicit policy (legal moves) & 67.0 & 71.4 \\
\bottomrule
\end{tabular*}

\vspace{6pt}
\begin{tabular*}{\linewidth}{@{\extracolsep{\fill}}lccc@{}}
\toprule
Confidence $\max_m q_m$ & Puzzles & Most likely & Sample \\
\midrule
below 0.4 & \phantom{0}91 & 39.6 & 30.4 \\
0.4--0.6  & 228 & 48.2 & 34.9 \\
0.6--0.8  & 243 & 63.8 & 51.3 \\
0.8 and above & 438 & 84.2 & 77.5 \\
\bottomrule
\end{tabular*}
\caption{\textbf{One distribution, several readouts.} Accuracy (\%) of the image model of \cref{tab:chess_recall_vs_play} on the first 1{,}000 test puzzles. \emph{Top:} measured, and expected if the samples were drawn from the implicit policy $q$. Single-sample rows are averaged over eight noise draws. \emph{Bottom:} the implicit policy's most likely move against a single 20-step sample, with the puzzles split by the policy's confidence.}
\label{tab:supp-readouts}
\end{table}

\noindent\textbf{Samples are draws from the distribution.} How often the eight samples choose a move follows its probability under $q$ (correlation 0.92), and $q$ predicts the accuracy of a single sample and of a vote over eight (\cref{tab:supp-readouts}, top).

\noindent\textbf{The most likely move beats a sample.} The implicit policy and the two-step output read the most likely move, and both are more accurate than a 20-step sample (\cref{tab:supp-readouts}, top). The two-step output is also nearly deterministic: all eight two-step samples agree on 76.9\% of the 10{,}000 test puzzles, against 27.6\% at 20 steps. The gap to a single sample is largest at intermediate confidence (\cref{tab:supp-readouts}, bottom), as \cref{eq:sample-vs-mode} predicts.

\begin{figure*}[t]
\centering
\includegraphics[width=\linewidth]{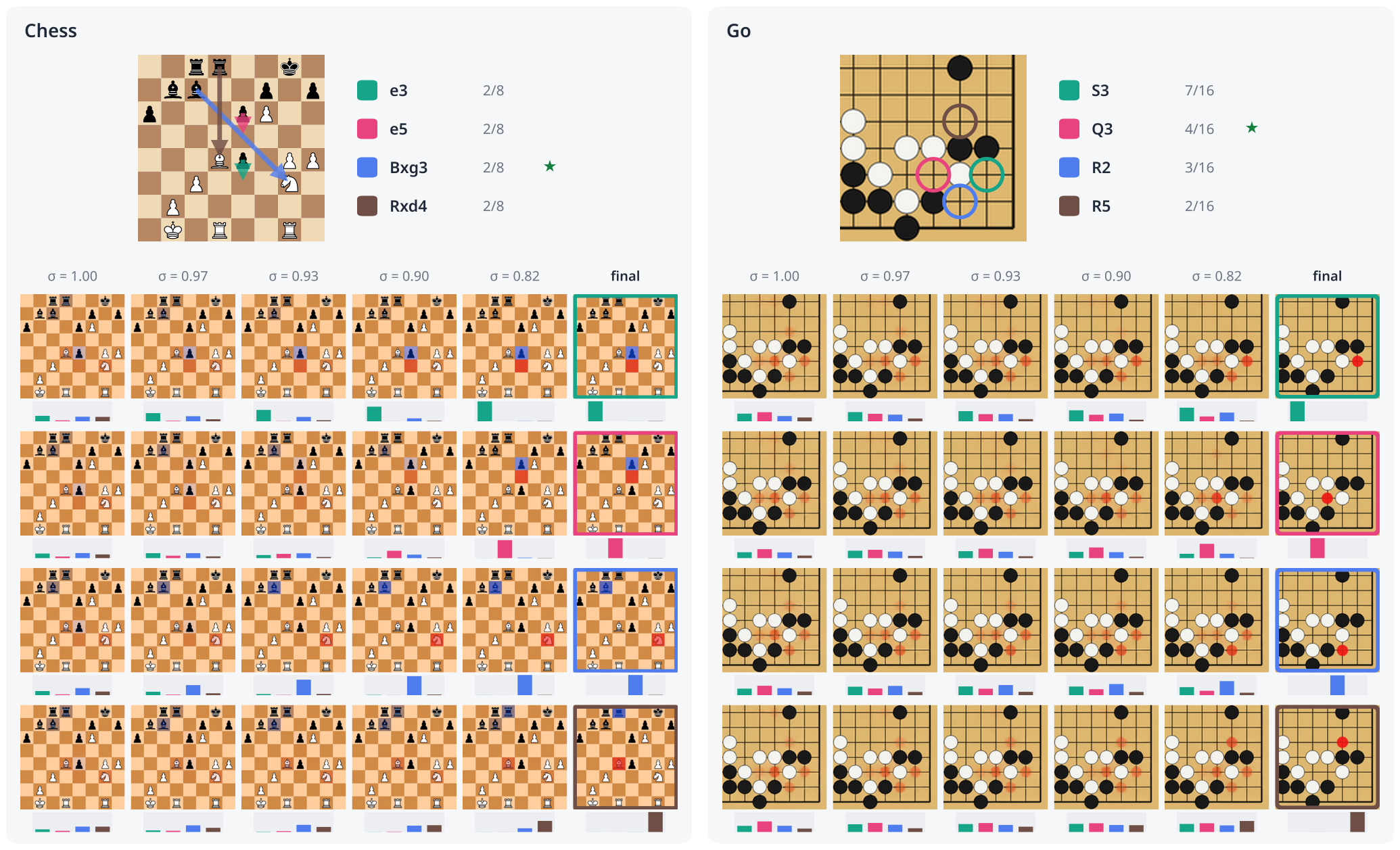}
\caption{\textbf{Decisions commit early.} Chess (left, Black to move) and Go (right, White to play, zoomed-in view). Each row follows one noise draw, and columns show decoded, unaltered clean-output estimates $\hat x_0$ at the labelled noise levels and the final 20-step output. Bars show candidate mark strengths. The outline colour identifies the final move, legend fractions give full-sample frequencies, and stars mark the correct moves.}
\label{fig:supp-denoise}
\end{figure*}

\begin{figure*}[t]
\centering
\includegraphics[width=\linewidth]{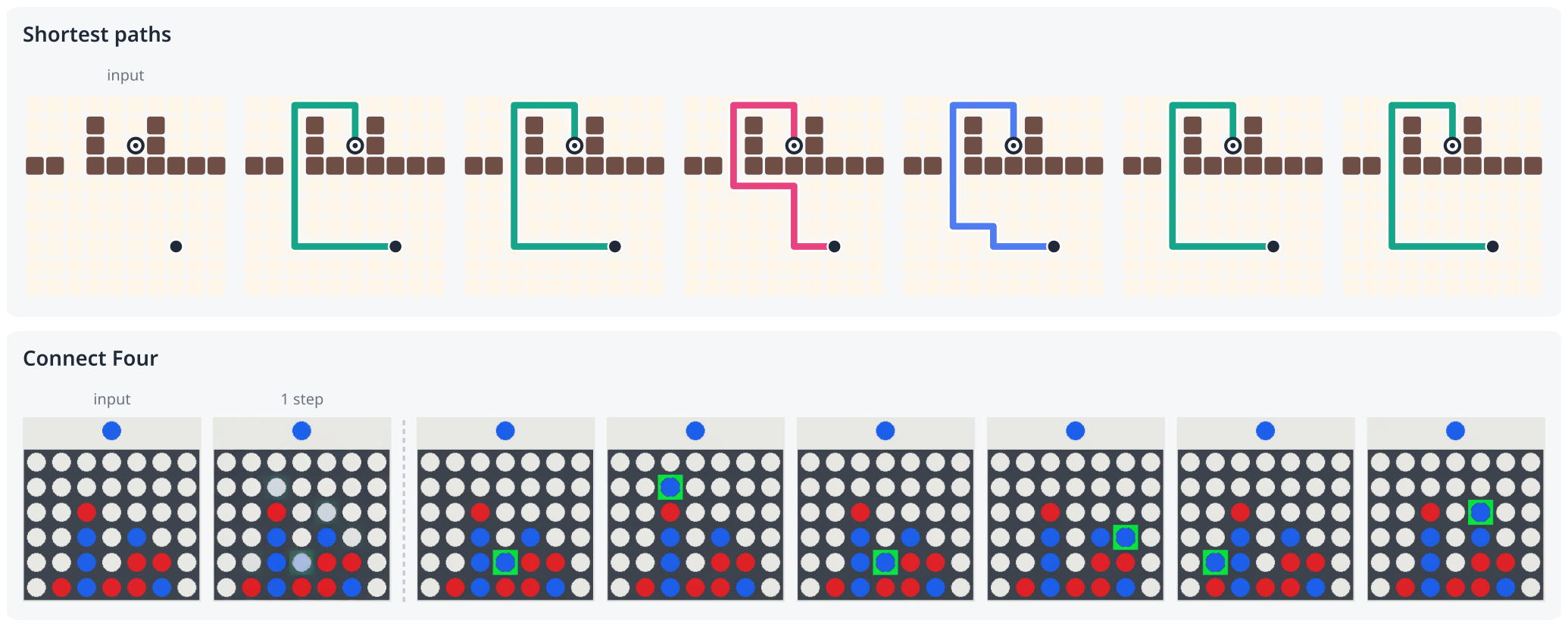}
\caption{\textbf{Different noise, different solutions.} \emph{Top:} Six routes on one map (dot: start, ring: goal). Matching colours indicate identical routes. \emph{Bottom:} A hard Connect Four position, its faint one-step output, and six unaltered 20-step samples to the right of the dashed line. The green square highlights the played disc. All shown routes are optimal, and all shown Connect Four moves preserve the win.}
\label{fig:supp-multisol}
\end{figure*}

\begin{table}[htb]
\centering\footnotesize
\begin{tabular*}{\linewidth}{@{\extracolsep{\fill}}lccc@{}}
\toprule
 & Candidates & Committed & Median \\
Model & at $\sigma{=}1$ & at step 1 & commit $\sigma$ \\
\midrule
Chess image, unguided & 3.7 & 47\% & 0.97 \\
Chess image, guided   & 2.0 & 65\% & 1.00 \\
Chess video           & 2.8 & 60\% & 1.00 \\
Path image            & 1.2 & 90\% & 1.00 \\
Path video            & 1.1 & 97\% & 1.00 \\
Terrain image         & 3.0 & 80\% & 1.00 \\
Terrain video         & 2.0 & 93\% & 1.00 \\
Go image, guided      & 55.1 & 37\% & 0.93 \\
\bottomrule
\end{tabular*}
\caption{Decisions commit early (validation items, chess models at one million puzzles).}
\label{tab:supp-dynamics}
\end{table}

\noindent\textbf{Decisions are fixed early.}
\label{app:denoising}
At every sampling step we decode the model's current estimate of the clean output and read it like a final output. To describe how many candidates it holds, we give every legal move the product of the blue strength at its origin and the red strength at its destination, normalise over the legal moves, and count the effective number of candidates ($\exp$ of the entropy). The legal moves are used only for this count. A decision \emph{commits} at the first step after which its leading move no longer changes. In every model and task, decisions commit at a median noise level of 0.93 or above (\cref{tab:supp-dynamics}): candidate moves differ in whole squares, far apart relative to the noise, so each sample is assigned to one of them almost immediately~\cite{biroli2024dynamical,raya2023spontaneous}, and the remaining steps only sharpen the image. Tasks differ in how many candidates the first estimate holds: few in pathfinding, where the target is nearly unique, and many in Go. \Cref{fig:supp-denoise} shows this progression for the chess image model of \cref{tab:chess_recall_vs_play} and the Go image model trained on one million positions. At $\sigma=1$, each noise draw contains several partially marked moves, and within a few steps it settles on one. Each row shows the first recorded noise draw ending on its move.

\begin{table}[htb]
\centering\footnotesize
\begin{tabular*}{\linewidth}{@{\extracolsep{\fill}}cccc@{}}
\toprule
Sampling & \multicolumn{2}{c}{Image model} & Video \\
\cmidrule(lr){2-3}
steps & unguided & guided & model \\
\midrule
1  & 38.0 & --   & \textbf{50.1} \\
2  & \textbf{38.3}  & 38.8 & 48.3 \\
4  & 34.8 & 41.5 & 47.4 \\
8  & 32.2 & \textbf{41.9} & 44.6 \\
15 & --   & --   & 42.5 \\
20 & \underline{31.4} & \underline{41.5} & -- \\
30 & --   & --   & \underline{41.9} \\
40 & 31.5 & --   & --   \\
\bottomrule
\end{tabular*}
\caption{Accuracy (\%) on 4{,}000 test puzzles against the number of sampling steps (models at one million puzzles). Underlined: default, bold: best.}
\label{tab:supp-steps}
\end{table}

\noindent\textbf{Fewer steps, better decisions.}
\label{app:steps}
\Cref{tab:supp-steps} varies the number of steps for the models at one million puzzles on the first 4{,}000 test puzzles. Without guidance, fewer steps give better decisions for both models, and every subset of puzzles improves as steps are removed (by solution depth, rating, number of legal moves and number of plausible alternatives), easy and single-solution puzzles improve more than hard and multi-solution ones. Where the steps are spent matters in the same direction: at a fixed number of steps, schedules that cross the high-noise range in one large step beat those that resolve it finely (image: 36.8 against 32.4\%, video: 47.8 against 42.8\%), because a large first step keeps the average and lands on its most likely move, while fine steps at high noise let the noise choose. Guidance is the exception: it replaces the conditional mean by an extrapolation, so Proposition~1 no longer applies, and fewer steps do not help. Reading the strongest mark without a minimum strength (\cref{app:eval}) matters here: the one-step output marks moves faintly, and a reader that requires a clear mark finds no move in 51.2\% of the video model's one-step outputs.


\subsection{Beyond chess}
\label{app:beyond}

\begin{table}[htb]
\centering\footnotesize
\setlength{\tabcolsep}{3pt}
\begin{tabular*}{\linewidth}{@{\extracolsep{\fill}}llccccc@{}}
\toprule
Task & Model & 1 & 2 & 4 & 8 & Full \\
\midrule
Chess$^\ast$ & image & \textbf{62.1} & 62.0 & 59.9 & -- & 57.3 \\
Chess$^\ast$ & video & \textbf{61.8} & 61.6 & 61.1 & -- & 55.6 \\
Go & image & \textbf{31.1} & \textbf{31.1} & 22.8 & 21.5 & 20.9 \\
Connect Four & image & \textbf{76.4} & 76.3 & 70.8 & 68.4 & 67.5 \\
Connect Four & video & \textbf{72.0} & 70.2 & 67.5 & 65.6 & 64.5 \\
Billiards & image & \textbf{28.0} & 27.9 & 24.0 & 20.2 & 18.6 \\
Billiards & video & \textbf{47.9} & 43.9 & 39.6 & 35.4 & 34.6 \\
\midrule
Terrain route & image & 87.3 & 87.2 & 86.5 & \textbf{87.7} & 87.5 \\
Terrain route & video & 62.6 & 71.6 & 80.4 & \textbf{82.8} & \textbf{82.8} \\
Shortest path & image & 89.0 & 89.1 & \textbf{89.3} & \textbf{89.3} & 88.8 \\
Shortest path & video & 83.0 & 88.9 & 94.0 & \textbf{94.7} & 94.6 \\
Detour or cross & image & 79.0 & 79.0 & 80.8 & 83.4 & \textbf{85.0} \\
\quad crossing is right & image & \textbf{90.8} & -- & 90.4 & 88.4 & 88.0 \\
\quad detour is right & image & 67.2 & -- & 71.2 & 78.4 & \textbf{82.0} \\
Detour or cross & video & 30.2 & 51.6 & 66.0 & 74.0 & \textbf{76.8} \\
\quad crossing is right & video & 41.2 & -- & 76.4 & 90.0 & \textbf{92.8} \\
\quad detour is right & video & 19.2 & -- & 55.6 & 58.0 & \textbf{60.8} \\
\bottomrule
\end{tabular*}
\caption{Accuracy (\%) against the number of sampling steps (full: 20 Euler steps for the image model, 30 UniPC steps for the video model). Chess$^\ast$: models of \cref{tab:chess_recall_vs_play}, 2{,}000 validation puzzles. Go: 2{,}000 test positions. Connect Four: value-preserving move on the 1{,}000 hard test positions. Billiards: red ball potted on 2{,}000 test layouts. Terrain: cheapest route on 1{,}000 test maps. Shortest path: optimal route on 1{,}000 held-out maps. Detour or cross: 500 probe maps, with indented rows splitting them by whether the cheapest route crosses the costly band or goes around it (250 maps each).}
\label{tab:supp-steps-tasks}\label{tab:supp-probe-split}
\end{table}

\noindent\textbf{Fewer steps on every task.} \Cref{tab:supp-steps-tasks} repeats the step comparison on every task and model. Wherever the output is a single decision, one or two steps are more accurate than full sampling: in chess, Go, Connect Four and billiards, for both models. This holds even in billiards, whose action is continuous: an average of several good shots could have been a poor shot, but it is not. Routes behave differently, and differently for the two models. The image model, which marks one move per call, is as accurate at one step on uniform maps (89.0 against 88.8\% optimal routes) and on the cheapest-route metric of terrain maps, but not on the probe maps, where the cheapest route may detour around costly terrain. The video model, which draws whole route segments, loses at few steps on every route task and needs about eight steps to match full sampling. Different noise can also produce different correct solutions (\cref{fig:supp-multisol}). The shortest-path map has 168 optimal routes of 17 moves. Six closed-loop runs of the image model, with one move per call and 20 steps per move, produce three distinct routes, and breadth-first search confirms that all six are optimal. In the hard Connect Four position, six of seven columns preserve Blue's win, and the first six 20-step samples choose five different value-preserving columns.

\begin{figure}[htb]
\centering
\includegraphics[width=\linewidth]{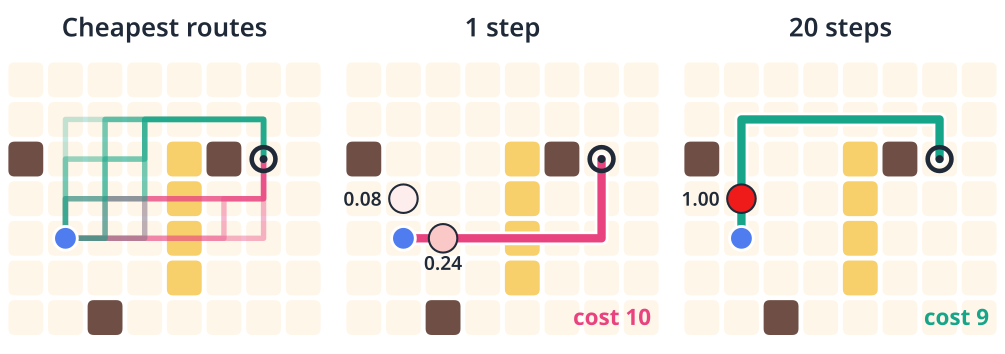}
\caption{\textbf{Why routes need more steps.} A terrain probe map where going around the costly band (yellow) is cheapest.
\emph{Left:} all cheapest routes around the band (green, cost 9) and the cheapest routes through it (pink, cost 10). Many of them share the first move towards the band.
\emph{Middle:} with one sampling step the image model outputs the average over these routes, so its first-move mark is faint and split (strength 0.24 towards the band, 0.08 around). The strongest mark is the shared cell, and the closed loop ends up crossing the band (cost 10).
\emph{Right:} with 20 steps the model commits to a single route, the mark is full (1.00) on the move around the band, and the route is the cheapest one (cost 9).
Mark strength is measured on the model's first output (0 = no mark, 1 = full mark).}
\label{fig:supp-route-steps}
\end{figure}

\noindent\textbf{Why routes need more steps.} The probe set separates two failure modes (\cref{tab:supp-probe-split}). For the image model, every error on a map that requires a detour is a route that crosses the costly band. One step is slightly more accurate than full sampling where crossing is right (90.8 against 88.0\%) and much less accurate where a detour is right (67.2 against 82.0\%). This is what reading the average of Proposition~1 predicts when the model is still uncertain at full noise: a crossing is a single straight route, so its probability sits on a few cells, while a detour can take many slightly different paths, so its probability is spread over many cells. The strongest cell of the average then lies on the crossing even when detours are more likely in total, whereas full sampling first commits to one route and then follows it (\cref{fig:supp-route-steps}). The video model shows no such bias. At few steps its clips stop before the goal (on 172 of the 250 detour maps at one step, against 8 at full sampling), because an average of clips is not a coherent clip. Among routes that reach the goal, the share of correct choices barely changes with the number of steps. We therefore use full sampling for all route tasks in the main results.

\begin{figure*}[!t]
\centering
\includegraphics[width=\linewidth]{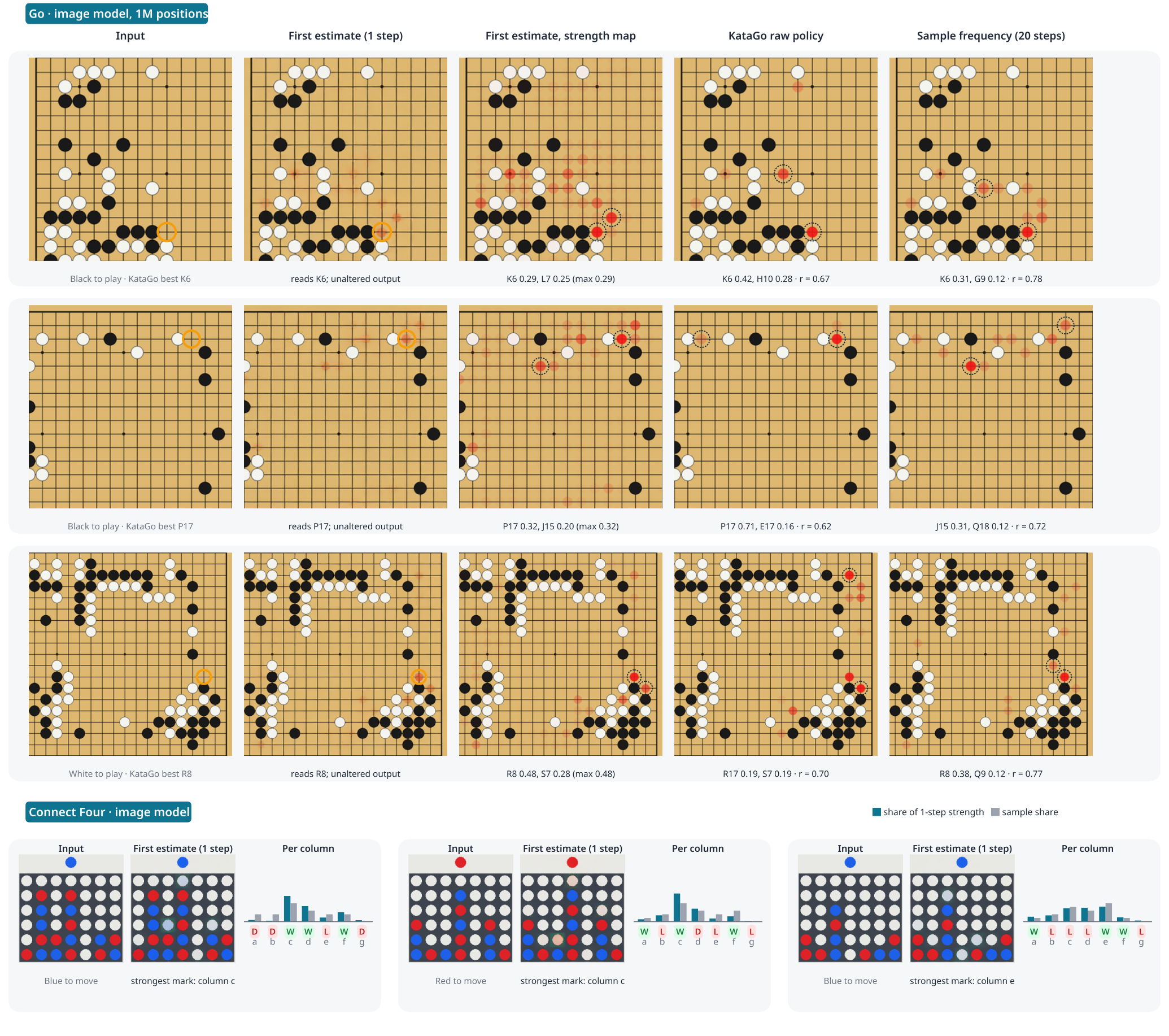}
\caption{\textbf{The first estimate on two other tasks.} \emph{Top,} Go: the position, the unaltered one-step output, the strength of each point in it, KataGo's policy network, and how often the model's 16 samples at 20 steps play each point (maps drawn as red discs scaled to their maximum). \emph{Bottom,} Connect Four: the position, the one-step output, which marks several columns, and per column its share of the one-step mark strength against the share of the samples that play it. W / D / L: the solver's value of each column, green: value-preserving. Rows: the positions with the most evenly split samples among those where the one-step decision is correct.}
\label{fig:supp-superposition-task}
\end{figure*}

\noindent\textbf{The first estimate is the model's own distribution.} Without an implicit policy, the theory can still be tested against the model's own samples: by \cref{eq:superposition}, an action should be marked in the one-step output as strongly as the full sampler chooses it. For Go and Connect Four we drew 16 full samples per position, on 1{,}000 positions each (\cref{tab:supp-first-estimate}). The mark strengths follow the sample frequencies on every model, the one-step decision agrees with the majority of the samples more often than a single sample does, and it is about as accurate as the vote over all 16 samples, at a sixteenth of the cost. In Go, the one-step output can also be compared with an external reference, the policy network of KataGo: its strongest point is KataGo's preferred move in 34.3\% of positions, against 20.0\% for a 20-step output, and its strength map agrees with KataGo's policy about twice as well (correlation 0.54 against 0.26). \Cref{fig:supp-superposition-task} shows examples.

\begin{table}[t]
\centering\footnotesize
\setlength{\tabcolsep}{3pt}
\begin{tabular*}{\linewidth}{@{\extracolsep{\fill}}lccccc@{}}
\toprule
 & Strength vs. & \multicolumn{2}{c}{Agrees with majority} & \multicolumn{2}{c}{Accuracy} \\
\cmidrule(lr){3-4}\cmidrule(l){5-6}
Model & frequency & 1 step & 1 sample & 1 step & vote \\
\midrule
Chess, image$^\dagger$ & 0.95 / 0.93 & 80.9 & 76.8 & 63.0 & 63.8 \\
Chess, video & -- & 81.8 & 70.6 & 61.0 & 63.3 \\
Go, image & 0.85 & 62.5 & 46.2 & 32.1 & 29.3 \\
Connect Four, image & 0.87 & 80.5 & 57.0 & 76.4 & 76.0 \\
Connect Four, video & 0.79 & 72.5 & 52.5 & 72.0 & 71.3 \\
\bottomrule
\end{tabular*}
\caption{\textbf{The one-step output against the model's own samples} (1{,}000 items each, with 16 samples at full steps and $^\dagger$8 for chess image). \emph{Strength vs.\ frequency:} Pearson correlation, over all items and actions, between an action's mark strength in the one-step output and the share of samples that choose it (chess: origin / destination squares, not measured for the video model). For chess these are 0.954 and 0.932, close to the correlations with the implicit policy in \cref{app:evidence} (0.954 and 0.929). \emph{Agrees with majority:} \% of items on which the one-step decision, or one full sample, equals the samples' most frequent action. \emph{Accuracy:} \% correct, one step against the vote over all samples.}
\label{tab:supp-first-estimate}
\end{table}

\subsection{What this means for inference}
\label{app:implications}

\noindent\textbf{Majority voting.}
\label{app:voting}
Samples from different noise can propose different moves, and a vote keeps the most frequent one. Voting therefore helps in proportion to how much the samples differ. At one million puzzles, at least one of eight samples is correct on 79.0\% of the puzzles for the video model, 70.0\% for the image model and 62.4\% with guidance, which concentrates the samples, and voting gains 10.4, 8.2 and 2.7 points (\cref{tab:supp-recall-1m}). Few-step samples are the extreme case: they all read the same average, so they barely differ, and voting adds little to them (\cref{tab:chess_recall_vs_play}).

\noindent\textbf{No search at inference.} Where iterative denoising performs optimisation, more steps improve the solution and harder instances need more of them~\cite{sun2023difusco}. We observe the opposite on both counts: accuracy falls with more steps, and harder positions gain less, not more, from them. At equal compute, samples are worth more than steps: on the 10{,}000 test puzzles, one step reaches 62.5\% with only one network evaluation, 20 steps 56.0\% with 20, and a vote over eight 20-step samples 63.8\% with 160. Look-ahead is acquired in training, as in chess models trained without search~\cite{ruoss2024amortized}, and extra computation at inference can only select among the model's own candidates. That is also where most of the remaining headroom lies: the correct move is among the eight 20-step samples for 85.4\% of the test puzzles and among the implicit policy's three most likely moves for 88.1\%, while no readout that does not use the answer exceeds 67.0\%. The same holds for whole puzzles: one or more of eight 20-step samples solves 69.0\% of them (\cref{app:whole}).

\noindent\textbf{Training at full noise only.} \Cref{eq:superposition} makes a prediction that can be tested directly. If one step returns the average over actions, then the full-noise part of the training loss, which is plain regression of the marked image, should be enough to produce the decision. We test this on Connect Four with the same backbone, LoRA, data, prompt and steps, but with every training example at $\sigma=1$. Read in one step, this model keeps the win on 83.5\% of the hard positions, against 76.4\% for the diffusion-trained model, as expected from a model that spends all of its training on the one noise level the readout uses. What it cannot provide is alternatives. Its outputs for eight noise draws agree on 98.4\% of positions, so they contain a value-preserving move on only 83.9\%, whereas eight 20-step samples of the diffusion-trained model, which agree on 3.7\%, contain one on 94.4\%. The one-step decision therefore comes from the full-noise term of the objective, as \cref{eq:superposition} states, and training across the other noise levels is what gives the same model a distribution to draw candidates from (\cref{fig:supp-multisol}).

\noindent\textbf{Scope.} The derivation assumes an optimal model and marks that add up. In practice the marks blur and the image decoder is nonlinear, so two-step outputs are more often illegal (8.6\%) than the mispairing of \cref{app:why} explains. With guidance the model no longer predicts a conditional mean, and fewer steps do not help. The implicit policy is somewhat overconfident (expected 71.4\% against a measured 67.0\%), and we measured calibration on one model only. The gain of the one-step output over a single sample is larger where the samples disagree, on every model, but not significantly.

\section{Recall or play: details}
\label{app:recall}

\begin{table*}[!t]
\centering
\small
\setlength{\tabcolsep}{3.0pt}
\renewcommand{\arraystretch}{1.15}
\begin{tabular}{@{}l ccc cc ccc@{}}
\toprule
& \multicolumn{3}{c}{Accuracy on held-out puzzles (\%)}
& \multicolumn{2}{c}{Symmetric boards (0 = no change)}
& \multicolumn{3}{c}{One piece changed (\%)} \\
\cmidrule(lr){2-4} \cmidrule(lr){5-6} \cmidrule(l){7-9}
& All $\uparrow$ & Sacrifices $\uparrow$ & Far from
& Colour & Mirror
& Keeps move: & Keeps move: & Finds new \\
& & & training $\uparrow$
& flip &
& best stays $\uparrow$ & best changes $\downarrow$ & best move $\uparrow$ \\
\midrule
\multicolumn{9}{@{}l}{\textit{Ours, one million training puzzles}} \\
Image model                 & 31.8 & 30.9 & 26.3 & \textbf{0.0}  & $-4.0$ & 62.5 & \textbf{42.7} & 20.4 \\
\quad + majority vote       & 40.0 & 43.3 & 33.7 & $+0.2$ & $-2.4$ & 73.4 & 47.9 & 23.2 \\
\quad + few-step            & 38.4 & 38.7 & 31.8 & $+0.7$ & $-2.4$ & 89.8 & 52.3 & 24.0 \\
Image model + guidance      & 40.9 & 42.2 & 34.5 & $+0.5$ & $-1.6$ & 82.1 & 49.8 & 29.5 \\
\quad + majority vote       & 43.6 & \textbf{47.4} & 37.4 & $+1.1$ & $-1.8$ & 87.0 & 53.4 & 29.9 \\
Video model                 & 41.7 & 35.9 & 35.4 & $+0.2$ & $-2.9$ & 91.7 & 50.5 & 25.6 \\
\quad + majority vote       & \textbf{52.1} & 46.8 & 46.1 & $+0.4$ & $-3.6$ & 92.0 & 50.8 & 29.0 \\
\quad + one step            & 50.5 & 40.6 & 45.0 & $+0.7$ & \textbf{$-$1.0} & \textbf{94.5} & 48.1 & 31.4 \\
\quad + one step + majority vote & 52.0 & 42.8 & \textbf{46.4} & $+0.4$ & $-2.1$ & 94.2 & 47.8 & \textbf{32.5} \\
\midrule
\multicolumn{9}{@{}l}{\textit{Copying from the same one million puzzles}} \\
Nearest training position   & 15.3 & 13.8 & \phantom{0}6.9 & $+0.4$ & $-5.2$ & 33.1 & 30.4 & \phantom{0}6.0 \\
Majority of 16 nearest      & 23.6 & 22.1 & 15.4 & $+0.2$ & $-7.0$ & 53.8 & 49.7 & 10.3 \\
\midrule
\multicolumn{9}{@{}l}{\textit{Direct prediction, trained on the same one million puzzles}} \\
Qwen3-VL-8B (board image)   & 42.0 & 43.1 & 34.1 & $-0.5$ & $-5.5$ & 91.8 & 61.1 & 21.6 \\
Qwen3-8B (FEN text)         & 40.9 & 41.4 & 32.7 & $+0.4$ & $-7.1$ & 90.5 & 65.6 & 18.2 \\
\midrule
Random legal move           & \phantom{0}5.5 & \phantom{0}3.8 & \phantom{0}4.0 & -- & -- & -- & -- & -- \\
\bottomrule
\end{tabular}
\caption{Recall-or-play results at one million puzzles, using the columns of \cref{tab:chess_recall_vs_play}. ``Guidance'' denotes contrastive guidance (\cref{app:guidance}), and the unguided image rows use the same mixture-trained model with $w=0$. ``Few-step'' uses the image model's validation-selected two steps. The 95\% confidence intervals are at most $\pm1.0$ points for all puzzles, $\pm3.1$ for sacrifices, $\pm2.0$ far from training, $\pm2.4$ for symmetry differences, and $\pm3.8$ for single-piece changes.}
\label{tab:supp-recall-1m}
\end{table*}

\noindent\textbf{Retrieval baselines.} To compare against memorisation, we look up, for each test position, the training position with the fewest differing squares and the same side to move, among all positions on the training lines the compared model was trained on: three million for \cref{tab:chess_recall_vs_play}, one million for \cref{tab:supp-recall-1m}. \emph{Nearest training position} plays that position's move on the test board, and \emph{majority of 16 nearest} plays the most common move among the 16 closest positions. Exact repeats are rare: only 17 of the 10{,}000 test positions occur on the three million training lines, each as a later position inside a solution line. None is the start position of a training puzzle, so none is a training input of either model. The distance to the nearest position also shows how differently the two strategies degrade. When a near-duplicate exists (at most 4 differing squares, 763 test puzzles), copying is right on 52.9\% of them, the majority of 16 on 55.4\% and our image model on 61.9\%. With 13 to 16 differing squares (1{,}757 puzzles), copying drops to 8.7\% but the image model solves 50.4\%.

\noindent\textbf{Symmetries.} Mirroring the board left to right, or swapping the colours together with flipping the board top to bottom, maps a puzzle onto an equivalent one whose answer is the transformed original answer. We apply each transformation to 2{,}000 test puzzles and compare accuracy before and after. Colour swaps leave accuracy unchanged for every model (all differences within 1.2 points). Mirroring lowers it by 0.5 to 4.0 points, so this symmetry is learned only approximately.

\noindent\textbf{One-piece changes.} The strongest test changes the position so that the best move may change. For each test puzzle we remove one piece other than a king, or move it to an empty neighbouring square, without touching the squares of the original answer. Stockfish~19 at depth 18 then decides whether the change is \emph{decisive}, meaning another move is now clearly best, or a \emph{control}, meaning the original move is still clearly best. A model that recalls a pattern should keep its move in both cases, while a model that evaluates the new position should keep it on control boards and abandon it on decisive ones. Our image model does the latter (\cref{tab:chess_recall_vs_play}). It keeps its move on 82.3\% of control boards but only on 35.3\% of decisive ones, and on the decisive boards it finds the new best move 39.5\% of the time, against 5.1\% for a random legal move. Retrieval barely reacts: the difference between control and decisive boards is 4.1 points for copying and 8.4 for the majority of 16 nearest positions, against 47.0 for our image model and 58.8 with few-step sampling. The video model at one million puzzles behaves in the same way (41.2 points, \cref{tab:supp-recall-1m}).

\begin{figure*}[!t]
\centering
\includegraphics[width=\linewidth]{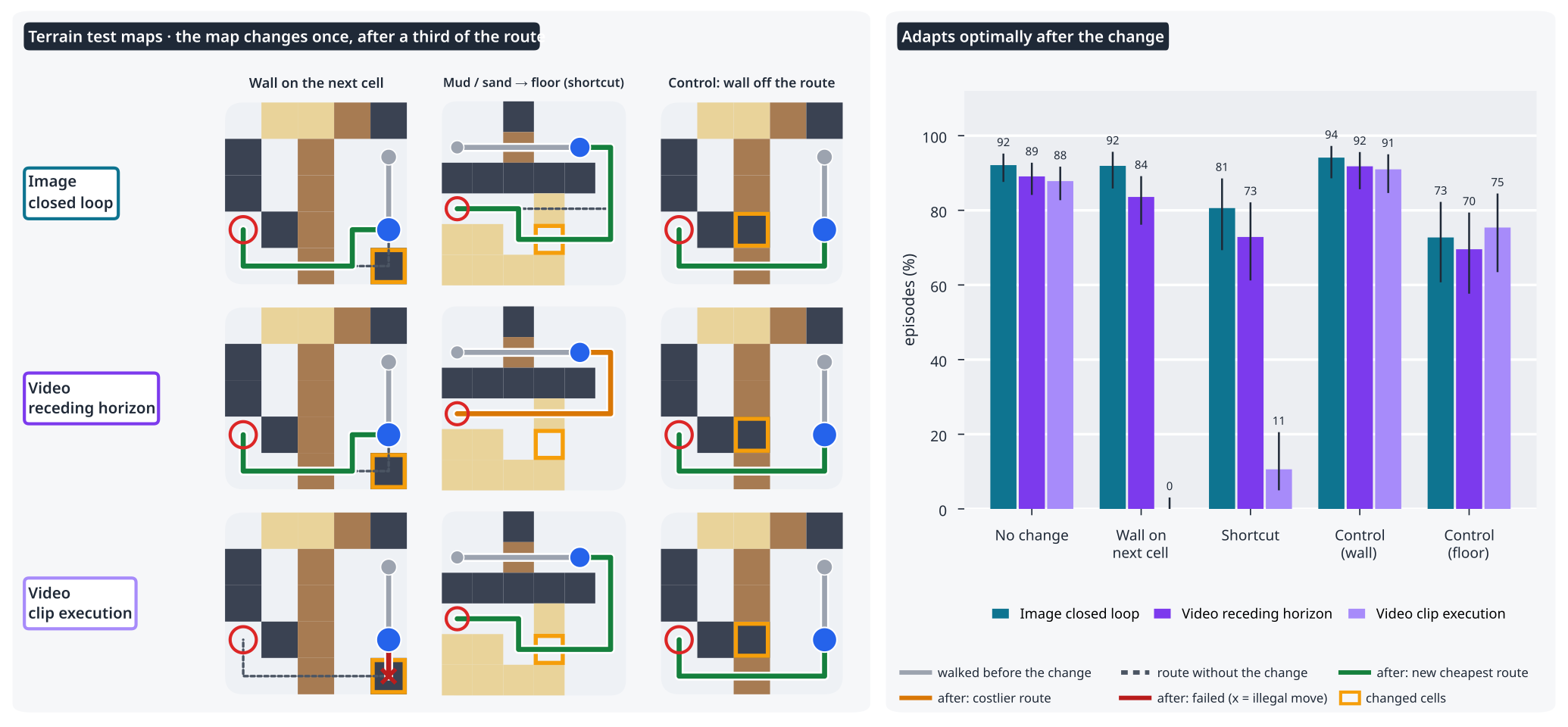}
\caption{\textbf{Changing a map mid-route.} After a third of the route, one change is made to the gold cells: a wall appears on the next cell, a cheaper shortcut opens through former mud or sand, or a wall appears away from the route as a control. \emph{Left:} examples for each change and model. Models that predict again after every move avoid the new wall, while a clip drawn before the change walks into it. \emph{Right:} the fraction of changed-map routes that are cheapest (95\% intervals).}
\label{fig:supp-midroute}
\end{figure*}
\noindent\textbf{Changing the map during a route.} We also changed the map once in the middle of a route. On the first 250 terrain test maps, each model walks from the start. After a third of its route, at least two moves before the goal, we either place a wall on the cell it would enter next, or turn mud or sand into floor so that a strictly cheaper route appears that the old route does not use. Changes of the same size away from every cheapest route serve as controls, and the episode continues on the changed map with the same sampling noise. \Cref{fig:supp-midroute} shows one map per change. When a wall appears on the next cell (left column), the image closed loop and the video receding horizon read the new state, turn away from the wall and follow a cheapest route of the changed map, while executing an already drawn video clip walks straight into the wall. When a shortcut opens (middle column), the closed loop takes it. In this example the receding horizon reaches the goal on a costlier route. A wall away from the route (right column) leaves all routes unchanged. Over all maps (\cref{tab:supp-midroute}), the two loops reach the goal after the wall in 98\% of episodes, never enter it, and end on a cheapest route about as often as without the change. When a shortcut opens, they take it in 82 and 80\% of episodes, against 10 and 19\% for their old routes. Executing whole clips does not adapt: the drawn moves walk into the new wall in every episode, and only 11\% of shortcut episodes end on the new cheapest route, because the model sees the change only at its next clip.
\begin{table}[htb]
\centering\footnotesize
\begin{tabular*}{\linewidth}{@{\extracolsep{\fill}}lccc@{}}
\toprule
 & Image, & Video, & Video, \\
Change during the route & closed loop & rec.\ horizon & clip \\
\midrule
None (all maps)            & 92.1 & 89.1 & 87.8 \\
\midrule
Wall on the next cell      & 91.9 & 83.6 & \phantom{0}0.0 \\
\quad \textit{same maps, no change} & \textit{91.2} & \textit{89.6} & \textit{91.0} \\
Shortcut opens             & 80.6 & 72.9 & 10.6 \\
\quad \textit{same maps, no change} & \textit{77.6} & \textit{71.4} & \textit{74.2} \\
\midrule
Control: wall off route    & 94.1 & 91.8 & 91.0 \\
\quad \textit{same maps, no change} & \textit{91.2} & \textit{89.6} & \textit{91.0} \\
Control: floor off route   & 72.7 & 69.6 & 75.4 \\
\quad \textit{same maps, no change} & \textit{77.3} & \textit{72.5} & \textit{75.4} \\
\bottomrule
\end{tabular*}
\caption{\textbf{Changing a map mid-route} on the first 250 terrain test maps. Entries give the percentage of episodes whose remaining route from the change point is a cheapest one on the changed map, among the episodes that reach that point. Italic rows show the same model and maps without a change. Because the measure starts at the change point, the first row is higher than the whole-route accuracy of \cref{tab:extended_tasks}. ``Clip'' executes each drawn clip in full.}
\label{tab:supp-midroute}
\end{table}

\noindent\textbf{One million puzzles, guidance and the video model.} \Cref{tab:supp-recall-1m} repeats \cref{tab:chess_recall_vs_play} at one million training puzzles, where both the image model (with and without contrastive guidance, \cref{app:guidance}) and the video model are available, together with retrieval from the same million puzzles. Every conclusion of the main table holds at this scale for all model variants. The models are far more accurate than copying, also far from training (26.3 to 46.4\% against 6.9 and 15.4\%), are unaffected by colour swaps, and keep their move on control boards much more often than on decisive ones, while retrieval barely tells the two apart. At equal data, the video model is the strongest of our single-sample predictors (41.7\%, against 39.9\% for the image model trained on the best-move prompt alone, and 31.8 and 40.9\% for the mixture-trained image model without and with guidance, \cref{app:guidance}), and it gains most from voting (52.1\%), because its samples are more diverse (\cref{app:voting}). Comparing the two scales shows how differently the two strategies use data: tripling the training set raises copying from 15.3 to 18.6\% (majority of 16: 23.6 to 27.0\%), but the image model trained on the best-move prompt alone from 39.9 to 56.0\%.

\section{Whole puzzles}
\label{app:whole}

Lichess counts a puzzle as solved only when every solver move is correct. We evaluate all 10{,}000 test puzzles under this rule. The first prediction uses the puzzle position, and later predictions use the true board after each recorded opponent reply, regardless of earlier predictions. This adds 15{,}679 later input positions. Neither model is trained to predict from them: the image model trains only on first positions, while the video model draws each line from its first position.

\begin{table}[htb]
\centering\footnotesize
\setlength{\tabcolsep}{3pt}
\begin{tabular*}{\linewidth}{@{\extracolsep{\fill}}llccc@{}}
\toprule
Model & Readout & First move & Later moves & Whole puzzle \\
\midrule
Image, one step & single & 62.5 & 62.8 & 38.3 \\
 & vote of 8 & 63.2 & 63.3 & 38.8 \\
 & any of 8 & 69.2 & 68.8 & 45.2 \\
Image, 20 steps & single & 56.0 & 57.0 & 31.1 \\
 & vote of 8 & 63.8 & 64.0 & 39.2 \\
 & any of 8 & \textbf{85.4} & \textbf{85.1} & \textbf{69.0} \\
Video, one step & single & 63.9 & 66.0 & 40.5 \\
 & vote of 4 & 64.4 & 66.7 & 41.0 \\
 & any of 4 & 70.2 & 72.2 & 47.9 \\
Video, 30 steps & single & 56.4 & 57.6 & 31.9 \\
 & vote of 4 & 62.8 & 63.8 & 38.5 \\
 & any of 4 & \underline{79.4} & \underline{80.9} & \underline{60.2} \\
\bottomrule
\end{tabular*}
\caption{Whole-puzzle results for the models in \cref{tab:chess_recall_vs_play} on 10{,}000 test puzzles (\%). Later moves are predicted from the true board after the recorded opponent reply. ``Any of $n$'' means that at least one of $n$ samples is correct.}
\label{tab:supp-whole}
\end{table}

One step gives the most accurate single readout for whole puzzles as well as first moves: solve rates are 38.3 and 40.5\%, versus 31.1 and 31.9\% with full sampling (\cref{tab:supp-whole}). Although trained only on first positions, the image model is similarly accurate on later and first moves (62.8 versus 62.5\%). Whole-puzzle success exceeds the rate predicted by independent moves (38.3 versus 33.1\%), indicating that difficulty tends to persist across a puzzle. The solve rate ranges from 89.8\% below rating 1000 to 3.8\% above 2500. Full sampling provides broader candidate coverage: at least one of eight samples at 20 steps solves 69.0\% of puzzles, compared with 45.2\% for eight one-step samples, which read nearly the same average. 

Grouped by the number of our moves in the recorded solution, the one-step video model solves 92.2, 49.4, 25.8 and 12.3\% of puzzles with one, two, three and four or more moves, respectively (image: 90.2, 47.0, 23.6 and 11.4\%). Whole-puzzle success falls further from three to four or more moves even though the video's first-move accuracy is nearly unchanged (57.4 versus 57.1\%).

\noindent\textbf{Are wrong moves alternative solutions?} A move outside the recorded solution may still be equally good. We compare each such move with the solution using Stockfish~19 at a fixed budget of 16M nodes per position. A move qualifies if it preserves a forced mate whenever the solution has one, or otherwise loses at most 5 points of win probability, the Lichess inaccuracy threshold. Alternatives are rare: another legal move qualifies in only 0.2\% of first positions and 0.7\% of later ones. Counting equally good model moves as correct raises per-move accuracy by at most 0.4 points and whole-puzzle solve rates from 38.3 to 38.8\% for the image model and from 40.5 to 40.9\% for video.

\section{How much lookahead do the puzzles need?}
\label{app:depth}

\begin{table}[htb]
\centering\footnotesize
\setlength{\tabcolsep}{3pt}
\begin{tabular*}{\linewidth}{@{\extracolsep{\fill}}lccccc@{}}
\toprule
Required depth & 1 & 2--3 & 4--5 & 6--9 & 10--30 \\
Puzzles & 7{,}748 & 1{,}246 & 537 & 317 & 145 \\
\midrule
Video, one step & 69.2 & 47.5 & 44.1 & 41.0 & 44.8 \\
Image, one step & 67.8 & 48.0 & 41.5 & 37.9 & 39.3 \\
\midrule
One-ply rule & 42.3 & 23.4 & 23.6 & 18.9 & 24.1 \\
Majority of 16 nearest & 29.2 & 19.4 & 20.3 & 18.9 & 15.9 \\
Random legal move & \phantom{0}5.6 & \phantom{0}4.9 & \phantom{0}4.8 & \phantom{0}4.8 & \phantom{0}5.1 \\
\bottomrule
\end{tabular*}
\caption{First-move accuracy (\%) of the models of \cref{tab:chess_recall_vs_play} by the shallowest Stockfish search depth at which the engine's best move is the puzzle answer. The one-ply rule plays a mate, otherwise the best capture, otherwise a check. On seven puzzles the engine does not play the answer at any depth up to 30, six of them mates in one with another equally good mate, and they are left out.}
\label{tab:supp-depth}
\end{table}

\noindent\textbf{Accuracy by required search depth.} Rating and solution length only approximate how much lookahead a puzzle requires. We therefore label each test puzzle with the shallowest depth at which Stockfish~19's best move is the puzzle answer (one thread, fixed hash, depth at most 30) and group first-move accuracy by this depth (\cref{tab:supp-depth}). The engine already plays the answer at depth 1 on 7{,}748 puzzles. Depth 1 is not a one-move rule, however: the engine still follows captures until the position is quiet and scores the result with a network trained on deeper searches, and a rule that plays a mate, the best capture or a check finds only 42.3\% of these answers, against 69.2\% for the one-step video model. On the 462 puzzles that need depth 6 or more, which are also much harder for human solvers (mean rating 2232 against 1649), the video model is right 42.2\% of the time and the image model 38.3\%, against 20.6\% for the one-ply rule, 18.0\% for the majority move of the 16 nearest training positions and 4.9\% for a random legal move. Accuracy drops once the answer is no longer the engine's immediate choice, but it does not keep falling as the required depth grows from 2 to 30, and much of the drop reflects puzzle rating: within rating bands, these deep puzzles are solved 2.4 points less often than the others by the video model and 8.3 by the image model. The 139 deep puzzles tagged as sacrifices behave alike (42.4\% for the video model, at most 22.3\% for any baseline). Connect Four, where an exact solver gives the number of moves to a forced win, shows the same pattern: the one-step image model keeps the win on 81.5, 84.9, 79.5 and 72.5\% of hard positions won in 1--3, 4--7, 8--11 and 12 or more of our moves, 14 to 21 points above the one-move heuristic in every group. The models thus stay well above shallow rules on positions that an engine solves only by searching deep, although they run no search.

\noindent\textbf{The one-ply rule and sacrifices.} On the three test sets of \cref{tab:chess_recall_vs_play}, the one-ply rule finds 38.0\% of the answers on all puzzles, 26.8\% on sacrifices and 33.6\% far from training, against 63.9, 55.6 and 59.1\% for the one-step video model. The sacrifice tag describes the whole solution, so we also check the moves themselves. By static exchange evaluation, our side gives up material somewhere in the solution on 857 of the 1{,}001 sacrifice puzzles and on the first move itself on 699. On these 699, the video and image models find the move 52.6 and 48.2\% of the time, against at most 26.6\% for the rule and the copying baselines.

\begin{table}[htb]
\centering\footnotesize
\begin{tabular*}{\linewidth}{@{\extracolsep{\fill}}lrcc@{}}
\toprule
Puzzle ends in & Puzzles & Image & Video \\
\midrule
Mate in 1 & 853 & 90.4 & 92.4 \\
Mate in 2 & 944 & 85.3 & 87.2 \\
Mate in 3 & 313 & 69.6 & 76.7 \\
Mate in 4 or more & 90 & 64.4 & 64.4 \\
No mate & 7{,}800 & 56.4 & 57.5 \\
\bottomrule
\end{tabular*}
\caption{One-step first-move accuracy (\%) of the models of \cref{tab:chess_recall_vs_play} by how the puzzle ends.}
\label{tab:supp-mate}
\end{table}

\noindent\textbf{Accuracy by how a puzzle ends.} First moves that start a forced mate are found more often than others, and less often as the mate gets longer (\cref{tab:supp-mate}). Even for mates in four or more moves, both models find the first move on 64.4\% of puzzles, above their accuracy on puzzles without a mate.

\section{Attention and occlusion}
\label{app:attention}

\Cref{fig:bishop-attention} in the main paper shows which pieces a chess decision depends on. We test this over many positions with the models at one million puzzles, on puzzles of two kinds: an enemy piece can be won, or the enemy king can be attacked. For the video model, attention from the decision frame to the piece that can be won is 1.13 times chance, against 0.86 for the other enemy pieces, and it stays higher after controlling for visual saliency (which squares hold pieces and which squares the move itself uses). In mate puzzles, the enemy king's zone receives 1.00 times chance against 0.89 in the other puzzles, whereas the pretrained model gives it 1.10 in both: fine-tuning makes attention depend on what the puzzle is about. Removing pieces confirms this causally: removing the piece that can be won lowers the strength of the chosen move by 0.61, removing another enemy piece by 0.02. For the image model, removing pieces shows the same dependence (0.67 against 0.05), but its attention does not. The piece that can be won receives more attention than the other pieces (1.15 against 1.05 times chance), but the pretrained model shows the same contrast, and attention to the king's zone does not differ between the two kinds of puzzle. The image model's decisions thus depend on the right pieces, but its attention maps do not show it.

Pathfinding allows a cleaner test because relevance is easy to define. Attention from the decision to the cells of the chosen route is above chance and above matched off-route cells for both video models (ratios 1.10 to 1.14), and the effect comes from fine-tuning. For the image models, attention to the route is at chance. On terrain maps, all three models attend more to the costly cells their route avoids than to matched floor cells, which shows that they look at the obstacles that shape the route. The counterfactual tests of \cref{app:recall} show the causal side: a wall on the chosen cell, or removing the terrain the route avoids, changes the decision, while the same change elsewhere does not.

\section{Probing for lookahead}
\label{app:probes}

The implicit policy shows that the image model distinguishes good moves from bad ones. Because move quality often depends on what follows, we ask whether the model represents future moves internally even though training asks it to draw only its own move.

\noindent\textbf{Features.} We run the unguided image model at one million puzzles exactly as at evaluation time and record the output of transformer blocks 8, 16, 24 and 32 (of 32) at sampling steps 0, 4 and 10 ($\sigma$ = 1.00, 0.86 and 0.60). Each board square covers $2{\times}2$ image tokens, and we average them into one 4{,}096-dimensional vector per square.

\noindent\textbf{Targets.} The puzzle's solution gives three sets of squares: those of our first move, which the model draws, of the opponent's reply and of our follow-up. For the reply we leave out the squares of the first move, and for the follow-up those of the first two moves. Otherwise a probe could succeed simply by finding the painted mark, for example when the reply recaptures on the destination square of our move.

\noindent\textbf{Probe and selection.} The probe is a single logistic-regression layer, shared by all squares, that maps a square's standardised vector to the probability that the square belongs to the target move. For each position, it ranks the candidate squares, and we score this ranking by its average precision. Probes are trained on 20{,}000 puzzles from training rows the model never saw. The block, the noise level and the regularisation are chosen on 1{,}834 validation puzzles, and only the chosen probe is evaluated, once, on 3{,}679 test puzzles.

\noindent\textbf{Baselines.} A probe could also succeed because the reply is easy to guess. To rule this out, we train the same probe on four other inputs. The first describes the board: for each square, the piece on every square of a $15{\times}15$ window around it, and the side to move. The second adds the true first move to this description, so that the probe knows what the model would play. The third uses the pretrained model's features, and the fourth pairs our model's features with the labels of a different position.

\begin{table}[htb]
\centering\footnotesize
\begin{tabular*}{\linewidth}{@{\extracolsep{\fill}}lcccccc@{}}
\toprule
Target & Chance & Board & +\,move & Pretr. & Ours & Gain \\
\midrule
Reply & 2.7 & 32.1 & 40.9 & 39.2 & \textbf{47.8} & +7.0 \\
Follow-up & 2.5 & 38.0 & 41.4 & 39.0 & \textbf{47.1} & +5.7 \\
\bottomrule
\end{tabular*}
\caption{Test-set average precision (\%) for the future-move probes. ``Gain'' is measured against the best baseline. The 95\% intervals are [+6.0, +7.9] for the reply and [+4.7, +6.7] for the follow-up.}
\label{tab:supp-probes}
\end{table}

\noindent\textbf{Results.} The fine-tuned model's features predict both future moves better than any baseline, including a board description that contains the true first move (\cref{tab:supp-probes}). Pretrained features perform about as well as the board description, indicating that fine-tuning adds the predictive information. Both moves are most readable at block 16, midway through the network, and are already readable at the first sampling step. For the reply, average precision is 47.8 at $\sigma = 1$ and 46.9 at $\sigma = 0.6$.

\section{The video model's imagined future}
\label{app:imagined}

The probes in \cref{app:probes} detect information about future moves inside the image model. The video model renders that future explicitly: after each decision, it draws the resulting board and the opponent's reply. We assess the accuracy of these imagined states and moves, and their relationship to the initial decision. Whether generating a continuation improves that decision is examined in \cref{app:horizon}.

\begin{table}[htb]
\centering\footnotesize
\begin{tabular*}{\linewidth}{@{\extracolsep{\fill}}lcccc@{}}
\toprule
 & Ply 1 & Ply 2 & Ply 3 & Ply 4 \\
 & (ours) & (reply) & (ours) & (reply) \\
\midrule
Correct & 41.7 & 20.1 & 12.2 & 2.0 \\
Correct, given correct prefix & -- & 40.9 & 39.5 & 17.1 \\
Imagined board exact & 99.1 & 98.0 & 96.3 & 93.6 \\
\bottomrule
\end{tabular*}
\caption{Accuracy of the video model's imagined line on 10{,}000 test puzzles (\%). Plies 2 to 4 are scored on the puzzles whose solution has that ply (9{,}141 for plies 2 and 3, and 4{,}594 for ply 4). A move counts as correct only when its raw drawing is legal on the imagined board and matches the puzzle move, so a clip that stops early counts as wrong. Imagined boards are scored after legal drawn moves.}
\label{tab:supp-plies}
\end{table}

\noindent\textbf{How good is the imagined line?} On 10{,}000 test puzzles, the video model trained on one million puzzles predicts the first move correctly 41.7\% of the time (\cref{tab:supp-plies}). Later moves are decoded from the model's own imagined boards, so errors accumulate. Given a correct first move, the imagined reply matches the puzzle in 40.9\% of cases. Given two correct moves, the model's follow-up matches in 39.5\%. Even when a reply differs from the puzzle, it can be reasonable: after a correct first move, 51.4\% are Stockfish's best replies and 63.6\% are within two points of win probability. The imagined boards are also accurate: after a legal drawn move, all 64 squares are correct in 93.6 to 99.1\% of cases, depending on the ply.

\noindent\textbf{The video clip is resolved in order.} The video model denoises every frame jointly, without an imposed temporal order, yet later boards tend to settle later. Relative to the board after the first move, the boards after the second, third, and fourth moves settle later in 54, 74, and 90\% of clips. This pattern follows the dependence between frames: a later board remains uncertain while earlier moves are unresolved. Indeed, when two first moves compete at high noise, the first imagined board blends their resulting positions, and as one move fades, so does its board. The model then resolves the next board, followed by later ones. This complements the early decision commitment in \cref{app:denoising} and shows that the clip is drawn as one dependent line of play and not as independent frames. It is a tendency, not a strict order: some later boards settle together with or before the first move.

\begin{figure*}[!t]
\centering
\includegraphics[width=\linewidth]{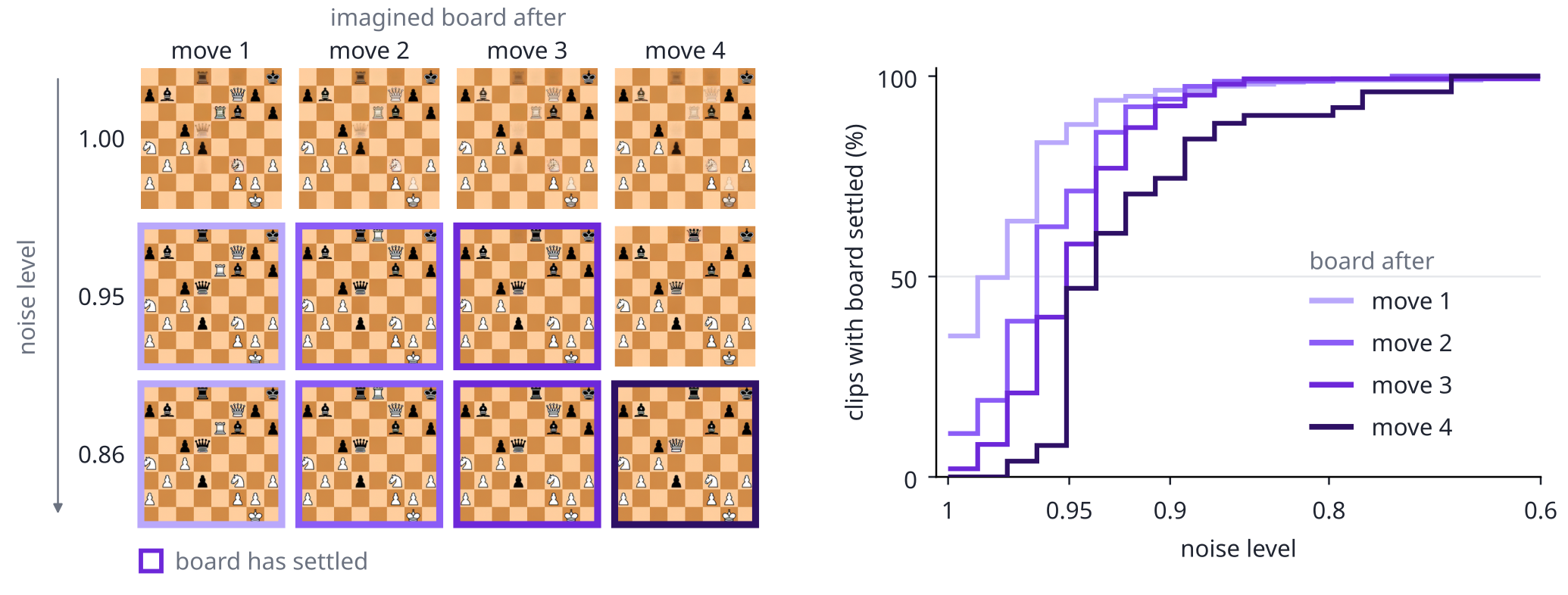}
\caption{\textbf{The clip is resolved in order.} \emph{Left:} one chess clip of the video model at three noise levels, shown as its unaltered estimates of the boards after moves 1 to 4. A frame marks a board that has settled. \emph{Right:} share of clips in which each board has settled, against the noise level. Relative to the board after move 1, the boards after moves 2, 3 and 4 settle later in 54, 74 and 90\% of clips.}
\label{fig:supp-settle}
\end{figure*}

\noindent\textbf{Wishful thinking.} After a wrong first move, the model tends to imagine a reply that makes its move look better than it is. We evaluate legal imagined replies on the true resulting board, ranking all opponent replies with Stockfish~19 at depth 12. The imagined reply is best in 51.4\% of puzzles after a correct first move but only 28.6\% after a wrong one, when the best reply is usually a refutation (Cochran--Mantel--Haenszel test stratified by how forced the position is, $p<10^{-100}$). Relative to the best reply, the imagined one leaves the moving side's win probability 35.9 points higher after a wrong move and 3.8 points higher after a correct move. Thus, an error often comes with an imagined continuation in which that error appears to work, rather than an independent check of the move.


\section{Prediction horizon}
\label{app:horizon}

We examine whether drawing a continuation improves the first move, what drives any gain, whether the model consults its imagined future at inference, and when the drawn moves should be executed. This extends the ``Learning from the continuation'' analysis in \cref{sec:experiments}.

\begin{table}[htb]
\centering\footnotesize
\begin{tabular*}{\linewidth}{@{\extracolsep{\fill}}llccc@{}}
\toprule
Task & Model & First move & With & $p$ \\
 & & only & continuation & \\
\midrule
Chess, 80k lines & video & \phantom{0}8.7 & 11.0 & $6\cdot10^{-13}$ \\
Chess, 400k lines & video & 21.4 & 24.6 & $5\cdot10^{-14}$ \\
Chess, 1M lines & video & 33.7 & 41.7 & $3\cdot10^{-64}$ \\
Terrain path & video & 82.5 & 95.0 & $3\cdot10^{-24}$ \\
Terrain path & image & 83.1 & 97.3 & $5\cdot10^{-31}$ \\
Connect Four, hard & video & 65.4 & 64.5 & 0.53 \\
\bottomrule
\end{tabular*}
\caption{First-move accuracy (\%) after training with or without a continuation, using the same data and training budget. Chess: 10{,}000 test puzzles. Terrain: 1{,}000 test maps, scored on the cost-optimal first move. Connect Four: value-preserving first move on 1{,}000 hard positions. $p$ values are exact McNemar tests.}
\label{tab:supp-horizon}
\end{table}

\noindent\textbf{Training with the continuation.} We compare targets that contain only the first decision and resulting state (\emph{first move only}) with targets containing the reference continuation of up to eight moves, including opponent replies (\cref{eq:horizon}). Data, training steps, learning rate, and seed are identical, and in chess both targets draw the green legal-destination dots (\cref{app:models}). Evaluation scores only the first decision. For the image model, the continuation target is a whole-route image ($H=L$). We execute its first move, render the resulting state, and call the model again, comparing it with a one-move model trained for the same 8k steps. At every training access the one-move model draws a position along the route and predicts the next move from it, so it sees the same route decisions that the whole-route model draws at once. Continuation training improves first-move accuracy in chess at every data size and in terrain pathfinding for both models (\cref{tab:supp-horizon}). Even after three times as much training (24k steps), the one-move terrain image model reaches 94.5\%, below the whole-route model's 97.3\% at 8k steps ($p=4\cdot10^{-4}$). Connect Four is an exception: the models are indistinguishable on the hard test set, and the first-move-only model is better on the standard set (69.1 versus 65.8\%, $p=2\cdot10^{-4}$). Because its training lines come from an exact solver, noisy continuations do not explain this limit.

For the video model trained on one million puzzles and evaluated with 30 sampling steps, continuation training improves first-move accuracy by 14.7, 8.3, 6.5 and 6.3 points on puzzles with one, two, three and four or more of our moves in the recorded solution. The gain spans all four groups and is largest on the one-move puzzles.

\noindent\textbf{The same decisions as separate examples.} A model trained on the continuation also sees more labelled decisions per example, so we compare it with controls that receive the same decisions as separate first-move examples. For the terrain image model, the one-move model already is such a control, since it trains on every position along the route: at the same compute it reaches 83.1\% against 97.3\%, and 94.5\% with three times the training. For the terrain video model, we train a first-move-only model on every route decision of the same segments as separate examples, with the same 5k steps. It reaches 85.3\%, against 82.5\% for the first move only and 95.0\% with the continuation ($p=4\cdot10^{-18}$ against the continuation model), which closes about a fifth of the gap. In chess, with 80{,}000 lines, the same control at equal compute reaches 6.3\%, below both the first move only (8.7\%) and the continuation (11.0\%, $p=4\cdot10^{-54}$). When its training is extended 2.3 times, so that every decision is seen once, it reaches 12.0\%, one point above the continuation model. The delayed-cue fork below gives the same answer in a controlled setting. At the same compute, drawing the continuation is therefore a more efficient training signal than the same decisions given separately. Separate examples need 2.3 times the training to match it in chess, and do not match it with three times the training on terrain.

\begin{figure*}[t]
\centering
\includegraphics[width=\linewidth]{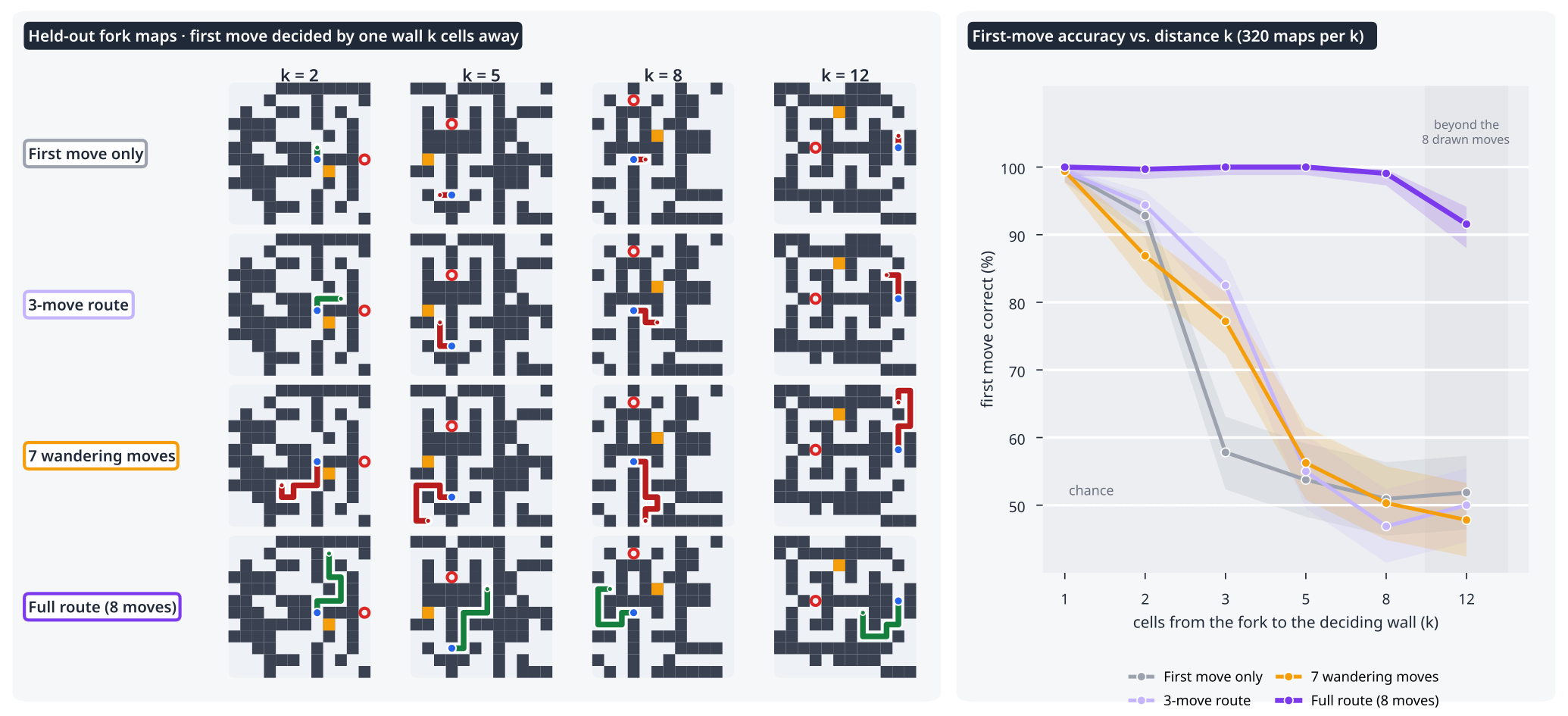}
\caption{\textbf{Delayed-cue fork.} \emph{Left:} held-out maps where a single wall $k$ cells away (orange) determines the first move, together with each of the four models' drawn routes. \emph{Right:} first-move accuracy against $k$ for the four training targets (320 test maps per $k$, 95\% intervals).}
\label{fig:supp-fork}
\end{figure*}
\noindent\textbf{Why the continuation helps: a delayed-cue fork.} We isolate the role of future supervision with maps containing two equal-length, mirror-image corridors. A single wall $k\in\{1,2,3,5,8,12\}$ cells down one corridor determines the first move. Every map has a twin with the wall in the other corridor, so ignoring it yields chance accuracy (50\%). Four video models train on the same 40{,}000 maps, steps, and seed. Their targets differ only after the correct first move: no continuation, the next eight route moves, the next three route moves, or seven moves that wander off-route. We score first moves on 320 held-out maps per $k$ (\cref{fig:supp-fork}). With this budget, the first-move-only model responds to walls one and two cells away (99.4 and 92.8\%), falls to 57.8\% at three cells and is at chance from five cells onward. With the route target, accuracy is 99 to 100\% through eight cells and 91.6\% at twelve, beyond the drawn continuation. Three route moves help through three cells, while the equally long wandering continuation helps at three but not from five. The gain therefore comes from learning to follow the route toward the decisive wall, rather than clip length alone.

\begin{figure}[htb]
\centering
\includegraphics[width=\linewidth]{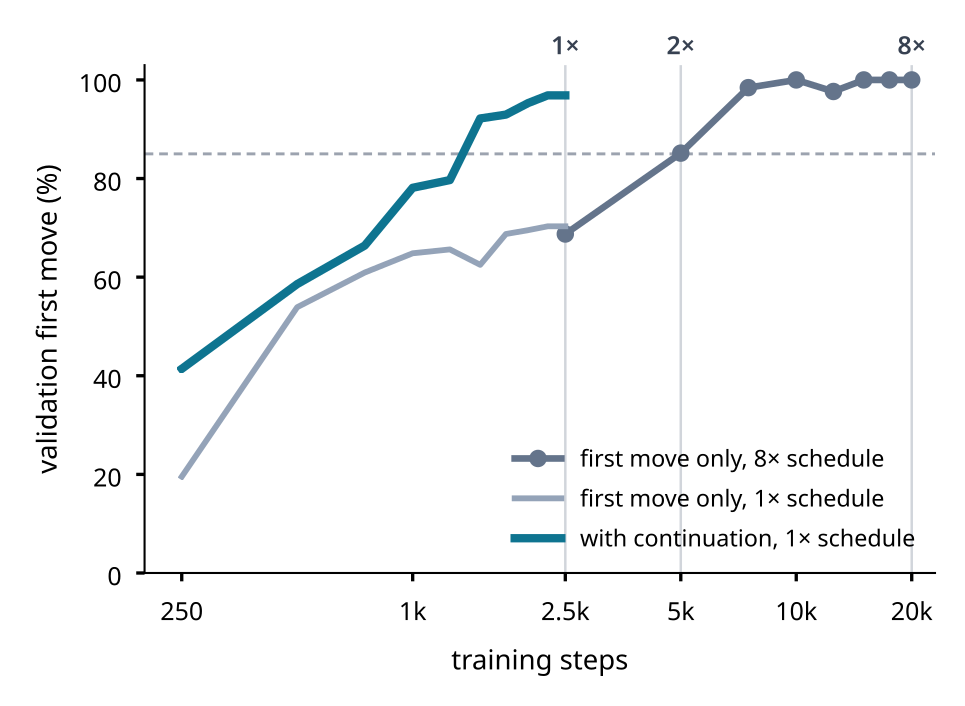}
\caption{\textbf{Learning the delayed-cue fork.} Validation first-move accuracy during training. Vertical lines mark one, two and eight times the training budget of \cref{fig:supp-fork}. With the continuation the model passes 85\% within 1{,}500 steps. Trained on the first move alone it needs 5{,}000 steps with a longer schedule and never exceeds 70\% with the original one.}
\label{fig:supp-fork-steps}
\end{figure}

\noindent\textbf{More labels or more training?} Two controls separate the continuation from what comes with it. First, the continuation supervises eight decisions per map instead of one. A first-move-only model trained on every position along the same eight-move routes as separate examples, with the same 2{,}500 steps, stays at chance for walls three to eight cells away (53.4\%, against 54.2\% for the first move only and 99.7\% with the continuation). Only one in eight of its examples shows the fork decision, so it is also weaker at near walls (43.1\% at two cells). With eight times the steps, when it has seen each fork as often as the first-move model, it reaches 94.2\% but remains below the continuation model at distant walls (85.3 against 99.1\% at eight cells and 70.0 against 91.6\% at twelve). Second, the first-move target can learn the rule, but it needs more training. With twice the steps on the same maps it reaches 84.7\% at three to eight cells and 70.9\% at twelve, still below the continuation model at every distance, and with eight times the steps it solves every test fork (\cref{fig:supp-fork-steps}). The reach of the continuation is also bounded by what is drawn: with the wall 10, 14 and 16 cells away, beyond the eight drawn moves, its accuracy falls to 96, 87 and 71\%. This is the pattern expected from intermediate supervision. A target that draws each step toward the cue lets every frame be supervised by evidence next to it, so a long-range dependency is learned as a chain of local ones, which makes composite rules learnable from far fewer examples~\cite{wies2022sub,gloeckle2024better}. The continuation thus lowers the amount of training a distant cue requires rather than making learnable what otherwise is not.

\begin{figure*}[!t]
\centering
\includegraphics[width=\linewidth]{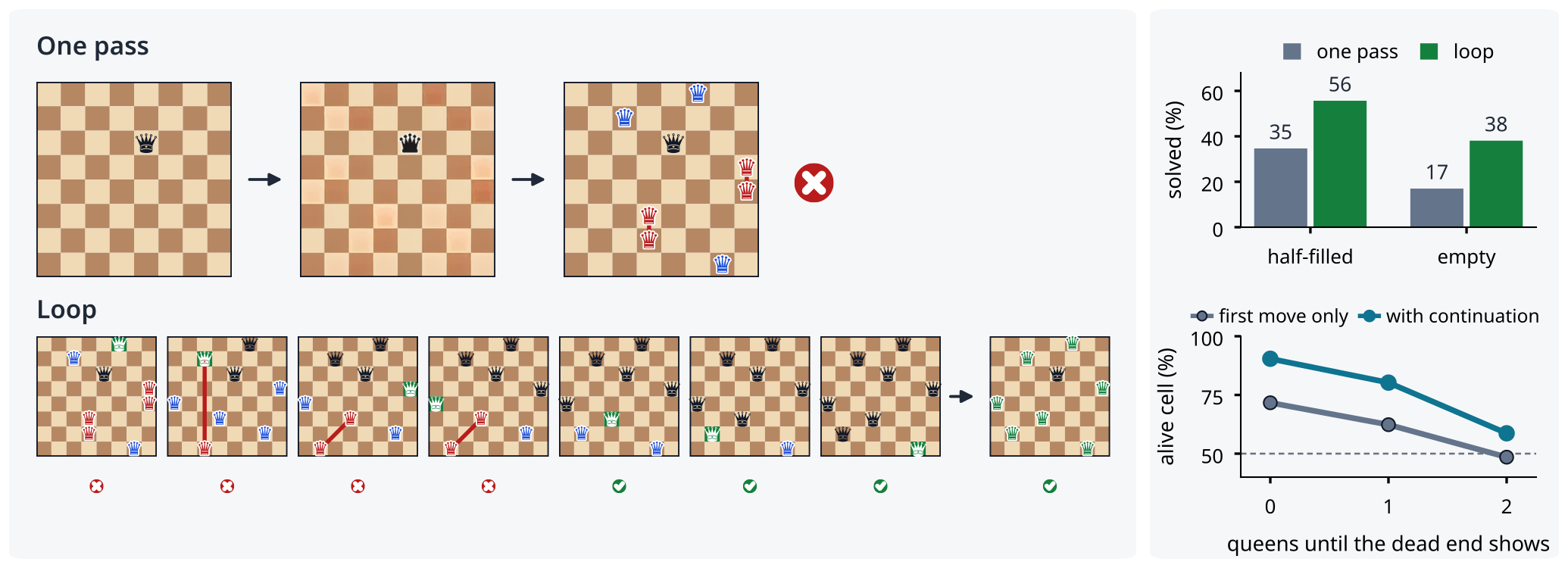}
\caption{\textbf{A wrong drawing can still guide the right move.} \emph{Left:} On a held-out board, one-pass completion fails, but executing the first queen and redrawing succeeds. Dark queens are already placed on the board, coloured queens are drawn, green marks the move taken, and crosses/checks mark completion validity. \emph{Right:} Solve rates for one pass and the loop (top), and first-move accuracy by dead-end depth (bottom, dashed line is chance).}
\label{fig:supp-nqueens}
\end{figure*}

\noindent\textbf{N-queens: a horizon measured in moves.} The delayed-cue fork measures distance to a decisive wall. N-queens lets us measure how many further placements are needed before a wrong first move reaches a dead end. We construct a mirror-balanced test of 3{,}606 boards on which the next row has exactly two candidate cells: one permits a complete solution, while the other reaches a dead end after further placements. Depths 0, 1 and 2 each have 600 mirror pairs (1{,}200 boards), while depth 3 has only three pairs and is reported descriptively below. Every board appears with its left--right mirror, so a choice that ignores the future scores 50\%. With first-move-only training, accuracy is 71.7, 62.3 and 48.5\% at depths 0, 1 and 2 (\cref{fig:supp-nqueens}, bottom right). At depth 3 it chooses the completable cell on 4 of 6 boards, too few for a comparable estimate. Taking 20 sampling steps instead of the validation-selected one does not extend the measured horizon (58.5, 53.6 and 46.6\% at depths 0 to 2).

\noindent\textbf{Training on the continuation.} In the \emph{with continuation} condition, the training target draws every missing queen, and in \emph{first move only} it draws just the next queen. Evaluation reads only the first move in both conditions. With the same training boards, recipe and seed, continuation training raises first-move accuracy to 90.4, 80.2 and 58.7\% at dead-end depths 0, 1 and 2 (\cref{fig:supp-nqueens}, bottom right). The depth-2 gain is 10.2 points ($p<0.001$). It also improves the first move on held-out solution families (92.6 versus 85.2\%) and hard boards (82.0 versus 62.0\%). Thus, future supervision helps the image model respond to consequences beyond the next placement, as it helps the video model respond to a distant wall in the fork task.

\noindent\textbf{Using the drawn solution.} We evaluate the model trained with continuation in two ways: draw a complete board once, or execute only its first queen and generate again from the resulting board. On 300 held-out half-filled boards, the closed loop solves 55.7\% with two sampling steps per call, against 34.7\% for the best one-pass result (20 steps). Across 512 empty-board trials, it solves 38.1\% against 17.0\% for the best one-pass result (50 steps, \cref{fig:supp-nqueens}, top right). On the example at left, the first four loop drawings are invalid as complete solutions, yet each supplies a completable next queen. Subsequent drawings become valid and the loop solves the board. The comparison changes how the drawing is used as well as the sampling schedule. It shows that the model's continuation is more useful when its first decision can be applied and revised than when the entire initial drawing is taken as the answer.

\noindent\textbf{Changing the board.} In a separate intervention, the environment adds a queen after the first move that contradicts the model's initially drawn solution. At two steps per call, a plan-breaking change is possible on 145 of the 300 half-filled boards. The loop solves 55.2\% of these changed episodes, against 48.3\% on the same boards without a change. A control adds the queen where the model had planned it. The plan-breaking condition outperforms this control on half-filled boards at two steps (55.2 versus 42.1\%, $p=0.011$), but underperforms it from empty boards (43.2 versus 51.9\%, $p=0.0007$). At 20 steps, the control comparisons are 46.3 versus 43.4\% on half-filled boards and 43.5 versus 40.8\% from empty boards, and neither difference is significant. Against no change, the plan-breaking condition is no worse numerically in the same four settings (55.2 versus 48.3, 43.2 versus 38.5, 46.3 versus 44.1, and 43.5 versus 40.8\%). These results show that the loop can continue from an unexpected state without an observed average penalty, but not that breaking its plan helps.

\noindent\textbf{The continuation is not read at inference.} If the model consulted its generated continuation to choose the first move, changing that continuation should change the decision. During video generation, we fix the opponent-reply frames to show either the engine's refutation of the model's first move or a weak reply, while keeping that first move visible. Under either condition, the first move changes on only 0.8\% of 1{,}000 validation puzzles (exact McNemar $p=1$). Together with the fork experiment, this points to a training benefit: predicting future states shapes the learned decision, but the drawn continuation is not consulted to make it.

\noindent\textbf{A parallel in language models.} Language models show a related effect under the name of multi-token prediction. Training a model to predict several future tokens at each position, instead of only the next one, improves sample efficiency, with the extra predictions serving as an auxiliary training task~\cite{gloeckle2024better}. The authors suggest that the gain comes from choice points, the positions whose outcome shapes what follows, which receive a stronger training signal when the future is predicted as well. A next best move is such a point. Our results show the same principle in a visual generator, where the predicted future consists of drawn states and opponent replies instead of tokens, and they allow two further tests: the drawn future can be changed at inference without changing the decision, and the same decisions given as separate examples do not replace it at the same compute.

\noindent\textbf{Executing the drawn moves.} In deterministic tasks, a predicted clip can be executed in full. On terrain maps, whole-clip execution finds a cheapest route on 82.8\% of test maps, compared with 83.0\% when predicting again after every move, and the probe-set rates are 76.8 and 78.8\%. On uniform maps, clips solve 94.6\% of test maps, versus 96.0\% for a receding horizon on the first 500. Each clip executes up to eight moves per call, rather than one, yielding 3.6 versus 19.4 seconds per executed move (\cref{tab:supp-infer-time}). Against an opponent, predicted replies may differ from actual ones. On the 1{,}411 chess puzzles where the video model correctly drew its first move and the opponent's reply and went on to draw a next move, that move is correct 41.2\% of the time, while a fresh prediction on the observed board after those same moves reaches 52.5\%. \Cref{tab:supp-plies} gives 39.5\% for the drawn move because it also counts the 62 clips that stop after the reply. Against the exact Connect Four solver, receding-horizon play converts 28.7\% of won starts, compared with 22.7\% when drawn moves are played regardless of the opponent's replies ($p=0.004$, 150 games).

\section{Data scaling}
\label{app:scaling}

\begin{table}[htb]
\centering\footnotesize
\begin{tabular*}{\linewidth}{@{\extracolsep{\fill}}lcccc@{}}
\toprule
Training & \multicolumn{2}{c}{Image, mixture} & Image, & Video \\
\cmidrule(lr){2-3}
examples & unguided & guided & best move & \\
\midrule
80k & -- & -- & -- & 12.7 \\
160k & \phantom{0}5.9 & 10.3 & -- & -- \\
400k & 14.6 & 19.9 & -- & 26.4 \\
1M & 30.4 & 39.4 & 39.7 & 41.6 \\
3M & -- & -- & 56.7 & 54.8 \\
\bottomrule
\end{tabular*}
\caption{Chess validation accuracy (\%) versus training-set size, averaged over three seeds (image: 2{,}000 puzzles, video: 1{,}000). ``Mixture'' uses the prompt mixture of \cref{app:guidance}, and ``best move'' uses the recipe of \cref{tab:chess_recall_vs_play}.}
\label{tab:supp-scaling}
\end{table}

We trained the chess models on increasing amounts of data with otherwise identical recipes, and evaluated each on validation puzzles with three sampling seeds (\cref{tab:supp-scaling}). Accuracy rises steadily for every model, and the gains do not slow down at the largest size, which suggests that data, not model capacity, limits these results. With the prompt mixture of \cref{app:guidance}, the image model's error rate falls roughly as a power law in the number of examples, with exponents of about 0.17 (unguided) and 0.21 (guided). The video model follows a similar curve (0.15). Each fit uses only three points, so the exponents are rough estimates. With the best-move prompt alone, tripling the data to three million puzzles raises validation accuracy from 39.7 to 56.7\%.

Two effects grow with data. First, contrastive guidance helps more as the model improves: its gain over unguided sampling is 4.4 points at 160k examples, 5.3 at 400k and 9.0 at 1M. Second, the advantage of imagining the full line over drawing a single decision grows from 2.3 to 3.2 and 8.0 test points at 80k, 400k and one million videos (\cref{app:horizon}). Go shows the same trend: the model trained on one million positions matches KataGo's move on 20.9\% of test positions, against 11.3\% when trained on 100{,}000 ($p < 10^{-20}$), and 27.9\% with majority voting.

\section{Contrastive guidance}
\label{app:guidance}

\noindent\textbf{Method.} A conditional diffusion model can be steered away from outputs it has learned to associate with a different prompt. For the one-move image model, the LoRA is additionally trained on two auxiliary prompts: one asking for any legal move, and one asking for a plausible but losing move $c^-$, labelled offline by the same engine that labels the reference moves (\cref{app:task-chess}). At inference, the velocity is steered away from the losing move,
\begin{equation}
  \tilde v = v_\theta(\cdot, c) + w\,\big(v_\theta(\cdot, c) - v_\theta(\cdot, c^-)\big),
  \label{eq:guidance}
\end{equation}
i.e.\ classifier-free guidance~\cite{ho2022classifier} with the unconditional branch replaced by a learned negative prompt. We use $w=3$. In Go, $c^-$ asks for a move that loses at least 20 points of win rate. No engine is queried at inference, and the cost is one extra network evaluation per denoising step (\cref{tab:supp-infer-time}). The horizon model always uses its plain sampler, since guidance needs a second pass over a whole video.

\noindent\textbf{Results.} Contrastive guidance improves the mixture-trained model, although the main results use unguided models. At one million chess puzzles, guidance raises single-sample accuracy from 31.8 to 40.9\% (\cref{tab:supp-recall-1m}). Much of this gain offsets the prompt mixture: training on the best-move prompt alone reaches 39.9\% without guidance, and training on three times as much data reaches 56.0\% (\cref{tab:chess_recall_vs_play}). Voting narrows the guidance gain from 9.1 to 3.6 points with eight samples (40.0 versus 43.6\%). Go shows a similar pattern at 100{,}000 positions: guidance raises single-sample accuracy from 9.3 to 14.0\%, versus 11.3\% for best-move-only training. With eight votes, the variants reach 14.1, 17.4, and 17.8\%. We therefore treat guidance as optional rather than using it in the main tables. Its advantage over unguided sampling grows with training data (\cref{app:scaling}).

\section{Does the result depend on the backbone?}
\label{app:backbones}

Qwen-Image uses Qwen3-VL to encode the prompt and input image, raising the possibility that the vision-language model solves the task while diffusion merely renders its answer. We test this on terrain pathfinding, where tile costs are hidden and must be inferred from examples. These image models draw an entire route in one image, unlike the closed-loop model of \cref{tab:extended_tasks}, which reaches the same 87.5\% on the test maps but 85.0\% on the probe maps. All variants use the same data, training steps, and learning rate.

\begin{table}[htb]
\centering\footnotesize
\begin{tabular*}{\linewidth}{@{\extracolsep{\fill}}llcc@{}}
\toprule
Model & Text encoder & Cheapest & Detour or \\
 & sees image? & route & cross \\
\midrule
Qwen-Image-2.1 & yes & 87.5 & 87.6 \\
Qwen-Image-2.1, blind encoder & no & 88.1 & 82.6 \\
FLUX.2 [klein] (4B) & no & 86.1 & 83.8 \\
Wan2.2 video & no & 82.8 & 76.8 \\
\midrule
Fewest steps (BFS) & -- & 59.2 & 50.0 \\
\bottomrule
\end{tabular*}
\caption{Terrain pathfinding with different backbones and text encoders (\% of test and probe maps). Image models draw a complete route in one image, and the video model draws and executes clips of up to eight moves.}
\label{tab:supp-backbones}
\end{table}

We first prevent Qwen-Image's text encoder from seeing the image, leaving the diffusion transformer as the sole visual input path. Cheapest-route accuracy remains 88.1 versus 87.5\% (exact McNemar $p = 0.59$). On probe maps it falls by 5.0 points, from 87.6 to 82.6\% ($p = 0.008$), so the encoder helps on difficult choices but is not necessary to infer the hidden costs: the blind model remains well above chance. We next use FLUX.2 [klein], a smaller 4B backbone with a text-only encoder. Its 86.1 and 83.8\% are not significantly different from Qwen-Image ($p = 0.29$ and $0.07$), and every model exceeds the terrain-blind planner (\cref{tab:supp-backbones}). The video model provides a third test because its text encoder is also text-only. On the same million chess puzzles, it reaches 41.7 versus 39.9\% for Qwen-Image on 10{,}000 test puzzles ($p = 0.001$). The latter reaches 40.9\% with contrastive guidance ($p=0.16$). Together, these results indicate that the visual decision capability is not confined to Qwen-Image or to an encoder that sees the image.

\section{Why generate the move? Direct-prediction baselines}
\label{app:baselines}

We compare models that predict moves directly from rendered images, or from text descriptions of the same positions for language models. They use the same puzzles or maps and validation sets, and are scored with the same rule-free readout: illegal answers count as wrong (\cref{tab:supp-baselines}). Strong symbolic chess transformers~\cite{ruoss2024amortized} use board encodings and chess-specific move vocabularies, so they are not directly applicable to the other tasks.

\begin{table}[htb]
\centering\footnotesize
\setlength{\tabcolsep}{3pt}
\begin{tabular*}{\linewidth}{@{\extracolsep{\fill}}lcc@{}}
\toprule
\multicolumn{3}{@{}l}{\textit{Chess, one million training puzzles}} \\
Model & Passes & Test (\%) \\
\midrule
Qwen3-VL-32B, zero-shot                 & 0 & \phantom{0}0.8 \\
Qwen3-VL-8B, fine-tuned                 & 1 & 42.0 \\
Qwen3-8B, text (FEN), fine-tuned        & 1 & 40.9 \\
DINOv2 + label output                   & 1 & 21.8 \\
DINOv2 + label output                   & 8 & 50.0 \\
DINOv2 + chess move head$^\dagger$      & 8 & \textbf{78.5} \\
\midrule
Ours, image   & 1 & 45.5 \\
Ours, video                    & 1 & \underline{50.5} \\
\bottomrule
\end{tabular*}

\vspace{2pt}
\begin{tabular*}{\linewidth}{@{\extracolsep{\fill}}lccc@{}}
\toprule
\multicolumn{4}{@{}l}{\textit{Delayed-cue fork, first move (\%)}} \\
 & \multicolumn{2}{c}{Training layout} & Right-angle \\
\cmidrule(lr){2-3}
Model & 1 pass & 12 passes & fork \\
\midrule
DINOv2, first move              & 67.2 & 100.0 & 58.0\\
DINOv2, first move + route      & 58.7 & 100.0 & 49.5 \\
\midrule
Ours, video, draws the route    & \textbf{98.4} & not run & \textbf{94.0} \\
\bottomrule
\end{tabular*}
\caption{\textbf{Direct-prediction baselines} trained on the same data. \emph{Top:} chess test accuracy by number of passes over the training puzzles, where one pass uses as many updates as our image model. The text model reads FEN, while the others read board images. Our rows use the number of sampling steps selected on validation (three for the image model, which is trained on the best-move prompt alone, and one for the video model). $^\dagger$The chess-specific output scores each of the $64{\times}64$ origin--destination pairs. \emph{Bottom:} delayed-cue fork. After one pass, classifiers use only nearby walls, and after twelve passes they solve every fork in the training layout.}
\label{tab:supp-baselines}
\end{table}

\noindent\textbf{Vision-language models.} Without fine-tuning, Qwen3-VL (8B and 32B) performs worse than a random legal move, whether given a board image or FEN text. Fine-tuned on the same one million puzzles to answer in UCI notation, the 8B model reaches 42.0\%, significantly below our one-step video model's 50.5\% (\cref{tab:supp-baselines}). The gap widens on tests that distinguish recall from play (\cref{tab:supp-recall-1m}). Far from training, accuracy is 34.1 versus 45.0\%. Mirroring costs $-5.5$ versus $-1.0$ points, and when one changed piece makes another move best, the model retains its old move in 61.1 versus 48.1\% of cases. These tests suggest that the visual generative model responds more reliably to task-relevant changes in the board than the vision-language model.

\noindent\textbf{A text-only language model.} We fine-tune the similarly sized Qwen3-8B on the same one million puzzles and with the same recipe as the vision-language model. It reads FEN and answers in UCI notation, using greedy decoding without reasoning before the answer. Its 40.9\% accuracy is significantly below the one-step video model (exact McNemar $p<10^{-60}$). The gap widens far from training (32.7 versus 45.0\%), mirroring costs $-7.1$ points, and on changed boards it keeps the old move in 65.6\% of cases where another move is best. With three million puzzles, it reaches 52.0\%, versus 62.5\% for the one-step image model on the same data. The two language models achieve similar accuracy and show similar failure patterns whether the input is an image or text. Compared with answering in text, drawing the move on the board yields higher accuracy with matched data, especially where the correct action depends on the current position rather than a familiar pattern.

\noindent\textbf{The output decides.} A generic transformer on frozen DINOv2-L features reaches 50.0\% after eight passes with a 4{,}096-way move output, or 48.1\% when the board is misaligned with its patches. Changing only the output to score moves at their origin and destination squares raises accuracy to 78.5\%, nearly identical to a head given the full board structure (78.6\%). The spatial output thus accounts for much of the classifier's strength. Drawing provides a spatial output across tasks without designing a separate head for each. After the same one pass and batch 8 used for our image model, the generic classifier reaches 21.8\%, versus 45.5\% for our model.

\noindent\textbf{Learning from each example.} With training to convergence, classifiers solve the delayed-cue fork (100\%) and produce optimal routes more often than our whole-route model. After one pass over the fork maps, matching our models' exposure, they respond to the wall only when it is one or two cells away (67.2\%, or 58.7\% with route supervision). Our route-drawing model reaches 98.4\%.

\noindent\textbf{Transfer to a new layout.} We also test forks whose corridors meet at a right angle, a layout absent from training. The converged classifiers reach 58.0 and 49.5\% and often move into a wall by choosing the direction favoured in the training layout. Our model retains 94.0\%. This suggests that the classifiers learned a layout-specific shortcut, whereas the route-drawing model transfers to the new geometry.

\begin{table}[t]
\centering\footnotesize
\setlength{\tabcolsep}{3pt}
\begin{tabular*}{\linewidth}{@{\extracolsep{\fill}}lcccc@{}}
\toprule
Task (test set) & Steps & Single-task & One model & Baseline \\
\midrule
Connect Four, keeps the win & 2 & 76.3 & 77.3 & 60.4 \\
Terrain, cheapest route & 1 & 87.3 & 83.4 & 59.2 \\
Terrain, cheapest route & 20 & 87.5 & 81.7 & 59.2 \\
Detour or cross & 1 & 79.0 & 79.0 & 50.0 \\
Detour or cross & 20 & 85.0 & 83.0 & 50.0 \\
\bottomrule
\end{tabular*}
\caption{\textbf{One model for two tasks} (\%). One image model is trained on Connect Four and terrain together, starting from the pretrained model. Single-task models use the same examples and training recipe. Baselines are the one-move heuristic for Connect Four, a terrain-blind planner for pathfinding, and chance for the detour-or-cross probe.}
\label{tab:supp-onemodel}
\end{table}

\noindent\textbf{One model for two tasks.} We jointly train one image model on Connect Four and terrain pathfinding, using the same prompt for both tasks. The model must identify the task from its input image, and each task contributes the same number of examples and uses the same recipe as its single-task counterpart (\cref{tab:supp-onemodel}). Joint training preserves Connect Four performance and the ability to decide whether to cross costly terrain, and stays within a few points of the single-task model on the cheapest route. Thus, one visual generator can support more than one task.

\section{Legality of the drawn actions}
\label{app:legality}

Because no model receives the rules, its raw drawing may be illegal, and all reported results count such actions as failures. Measured directly on the test sets, almost all raw drawings are legal. In chess, the image model trained on three million puzzles draws a legal move on 97.0\% of positions (88.3 to 88.7\% at one million puzzles), and the video model trained on one million on 93.4\%. After legal drawn moves, 97.3\% of imagined video boards are correct on all 64 squares, with 99.95\% of individual squares correct. In Go, the marked point is legal 99.8\% of the time, versus 43\% for the base model. Connect Four discs occupy a non-full column and the correct landing row in 100.0\% of image outputs and 99.6\% of video outputs, although gravity is never stated. For the image and video models respectively, routes are valid on 94.7 and 97.4\% of shortest-path maps and on 98.5 and 95.0\% of terrain maps, and 99.2 and 98.2\% of billiards outputs are readable shots.

\section{Runtime and memory}
\label{app:runtime}

\begin{table}[tb]
\centering\footnotesize
\begin{tabular*}{\linewidth}{@{\extracolsep{\fill}}llrr@{}}
\toprule
Training & Model & Wall-clock & GPU-hours \\
\midrule
Chess, 3M$^\ast$ & image / video & 93\,h / 79\,h & 1{,}487 / 1{,}268 \\
Chess, 1M & image / video & 51\,h / 49\,h & 407 / 389 \\
Go, 100k / 1M & image & 19\,h / 90\,h & 152 / 722 \\
Terrain path & image / video & 24\,h / 9\,h & 195 / 75 \\
Connect Four & image / video & 9\,h / 8\,h & 75 / 66 \\
Billiards & image / video & 32\,h / 22\,h & 260 / 172 \\
\bottomrule
\end{tabular*}

\vspace{5pt}
\begin{tabular*}{\linewidth}{@{\extracolsep{\fill}}lcr@{}}
\toprule
Inference & Steps & GPU-s / decision \\
\midrule
Chess image, unguided / guided & 20 & 3.1 / 6.1 \\
Chess image, one step & 1 & 0.7 \\
Chess video, full clip / one step & 30 / 1 & 8.9 / 2.2 \\
Go image, 384\,px, unguided / guided & 20 & 6.8 / 13.1 \\
\midrule
Terrain path, image closed loop & 20 & 3.3 per move \\
Terrain path, video & 30 & 3.6 per move \\
Terrain path, video, receding horizon & 30 & 19.4 per move \\
\bottomrule
\end{tabular*}
\caption{\textbf{Training and inference cost.} \emph{Top:} training on A5000 GPUs, including periodic validation. Each run uses eight GPUs, except $^\ast$ runs with sixteen GPUs across two nodes. \emph{Bottom:} measured inference cost, including model loading and decoding. Pathfinding entries give wall-clock seconds per executed move.}
\label{tab:supp-train-time}\label{tab:supp-infer-time}
\end{table}

Training uses nodes with eight NVIDIA RTX A5000 GPUs (24\,GB each). The image model and its text encoder do not fit together on one GPU, so each process uses two: one holds the diffusion transformer, the other Qwen3-VL and the VAE. At $256$ pixels and two examples per process, peak memory is 18.3\,GB on the transformer's GPU and 18.5\,GB on the encoder's. The video model fits on one GPU, peaking between 15 and 18\,GB depending on resolution. \Cref{tab:supp-train-time} reports training costs for the main runs.

Inference cost depends chiefly on the number of denoising steps and on whether guidance adds a second forward pass (\cref{tab:supp-infer-time}). One-step sampling (\cref{app:eval}) is the cheapest setting and, without guidance, also the more accurate one. Voting multiplies the cost by the number of samples. The execution protocol matters on move-by-move tasks: a video clip executes up to eight moves per call and solves a terrain map in about the time of the image model's closed loop, while the receding horizon generates a new clip for every executed move and is about five times slower.